\documentclass[a4paper,11pt]{article}
\pdfoutput=1 % to allow compilation in JCAP
\usepackage{jcappub}
\usepackage{amsmath}
\usepackage{bm}
\usepackage{graphicx}
\usepackage{enumitem}
\usepackage{geometry}
\usepackage{xspace}
\usepackage{pdflscape}
\usepackage{dsfont}

\usepackage{relsize}

\allowdisplaybreaks % allows long equations to be displayed over several pages

\makeatletter
\gdef\@fpheader{}
\g@addto@macro\bfseries{\boldmath}
\makeatother

\newcommand{\ie}{{i.e.~}}

\DeclareMathOperator{\erfc}{erfc}

\newcommand{\dd}{\mathrm{d}}
\newcommand{\ee}{e}

\newcommand{\sss}[1]{{\scriptscriptstyle{#1}}}

\newcommand{\uPl}{\mathrm{Pl}}

\newcommand{\usssPl}{\sss{\uPl}}

\newcommand{\Mp}{M_\usssPl}

\newcommand{\efolds}{$e$-folds}

\newcommand{\beq}{\begin{equation}}
\newcommand{\eeq}{\end{equation}}
\newcommand{\bea}{\begin{equation}\begin{aligned}}
\newcommand{\eea}{\end{aligned}\end{equation}}

\newlength{\wsingfig}
\newlength{\wdblefig}
\newlength{\wquadfig}
\newlength{\wtriplefig}
\newcommand{\Eq}[1]{Eq.~(\ref{#1})}
\newcommand{\Eqs}[1]{Eqs.~(\ref{#1})}
\newcommand{\Fig}[1]{Fig.~{\ref{#1}}}

\newcommand{\Ref}[1]{Ref.~{\cite{#1}}}
\newcommand{\Refs}[1]{Refs.~{\cite{#1}}}
\newcommand{\Sec}[1]{Sec.~\ref{#1}}
\newcommand{\Secs}[1]{Secs.~\ref{#1}}
\newcommand{\App}[1]{Appendix~\ref{#1}}

\newcommand{\deflen}[2]{%      
    \expandafter\newlength\csname #1\endcsname
    \expandafter\setlength\csname #1\endcsname{#2}%
}

\newcommand{\old}[1]{} % if not

\subheader{}

\title{Tunneling in Stochastic Inflation}

\author[a,b]{Mahdiyar~Noorbala,}
\author[c,d]{Vincent~Vennin,}
\author[d]{Hooshyar~Assadullahi,}
\author[b]{Hassan~Firouzjahi}
\author[d]{and David~Wands}

\affiliation[a]{Department of Physics, University of Tehran, Iran, P.O.~Box 14395-547}
\affiliation[b]{School of Astronomy, Institute for Research in Fundamental Sciences (IPM), Tehran, Iran, P.O.~Box 19395-5531}
\affiliation[c]{Laboratoire  Astroparticule  et  Cosmologie,  Universit\'e Denis  Diderot  Paris  7,  10  rue Alice Domon et L\'eonie Duquet, 75013 Paris, France}
\affiliation[d]{Institute of Cosmology \& Gravitation, University of Portsmouth, Dennis Sciama Building, Burnaby Road, Portsmouth, PO1 3FX, United Kingdom}

\emailAdd{mnoorbala@ut.ac.ir}
\emailAdd{vincent.vennin@apc.univ-paris7.fr}
\emailAdd{hooshyar.assadullahi@port.ac.uk}
\emailAdd{firouz@ipm.ir}
\emailAdd{david.wands@port.ac.uk}

\date{today}

\begin{document}
\sloppy

\abstract{The relative probability to decay towards different vacua during inflation is studied. The calculation is performed in single-field slow-roll potentials using the stochastic inflation formalism. Various situations are investigated, including falling from a local maximum of the potential and escaping from a local minimum. In the latter case, our result is consistent with that of Hawking and Moss, but is applicable to any potential. The decay rates are also computed, and the case of a generic potential with multiple minima and maxima is discussed. }

%\arxivnumber{18XX.XXXXX}
\maketitle

\section{Introduction}
The inflationary paradigm~\cite{Starobinsky:1980te, Sato:1980yn, Guth:1980zm, Linde:1981mu, Albrecht:1982wi, Linde:1983gd} is usually regarded as the most promising theory describing the early stages of the evolution of the universe. In addition to solving some of the hot big bang model problems, it offers a specific mechanism, namely quantum fluctuations~\cite{Mukhanov:1981xt, Hawking:1982cz,  Starobinsky:1982ee, Guth:1982ec, Vilenkin:1983xp, Bardeen:1983qw, Starobinsky:1979ty}, for generating the primordial cosmological perturbations that are observed on the cosmic microwave background and seed the late-time structures.

On super-Hubble scales, these quantum fluctuations exhibit a ``classical'' behaviour~\cite{Polarski:1995jg, Lesgourgues:1996jc, Kiefer:2008ku, Burgess:2014eoa, Martin:2015qta, Grain:2017dqa}, in the sense that most of their statistical properties can be well described by a background motion under the influence of a classical noise. For a scalar field, the long-wavelength modes can be incorporated into a coarse-grained field $\phi$, whose evolution is governed by the Langevin equation
\bea
\label{eq:Langevin}
\frac{\dd\phi}{\dd N} = -\frac{V'}{3H^2}+\frac{H}{2\pi}\xi\left(N\right)
\eea
in the slow-roll regime. In this expression, $N\equiv \ln a$ is the number of \efolds~where $a$ is the scale factor, $V'$ is the derivative of the potential $V$ with respect to the field value, $H\equiv \dot{a}/a$ is the Hubble factor where a dot denotes a derivative with respect to cosmic time, and $\xi$ is a white Gaussian noise with zero mean and unit variance. This noise accounts for the continuous entry of short-wavelength modes into the coarse-grained sector. This formalism is known as ``stochastic inflation''~\cite{Starobinsky:1982ee, Vilenkin:1983xp, Starobinsky:1986fx, Nambu:1987ef, Nambu:1988je, Kandrup:1988sc, Nakao:1988yi, Nambu:1989uf, Linde:1993xx, Starobinsky:1994bd}.
 
It can be used to compute the correlation functions of the fields~\cite{Starobinsky:1994bd, Finelli:2008zg, Finelli:2010sh, Garbrecht:2013coa, Garbrecht:2014dca, Prokopec:2015owa, Boyanovsky:2015jen, Boyanovsky:2015xoa, Boyanovsky:2015tba, Burgess:2015ajz, Prokopec:2007ak, Prokopec:2008gw, Tsamis:2005hd, Grain:2017dqa} and gives results in agreement with standard quantum field theoretic calculations. Combined with the $\delta N$ formalism~\cite{Starobinsky:1986fxa, Sasaki:1995aw, Wands:2000dp, Lyth:2005fi} into the so-called ``stochastic $\delta N$ formalism'', it also gives rise to a method to compute correlation functions of cosmological perturbations that incorporates quantum backreaction effects~\cite{Fujita:2013cna, Fujita:2014tja, Vennin:2015hra, Assadullahi:2016gkk, Vennin:2016wnk, Tada:2016pmk,  Pattison:2017mbe, Tokuda:2017fdh}.

In this paper, we show how stochastic inflation can be used to study another interesting problem, namely the calculation of the relative probabilities to decay towards different vacua in symmetry-broken potentials~\cite{Linde:1991sk, Tolley:2008qv}. In practice, we make use of the ``first passage time'' techniques developed in \Refs{Vennin:2015hra, Assadullahi:2016gkk, Vennin:2016wnk} where it is shown that if the inflaton field takes initial value $\phi_0$ between $\phi_-$ and $\phi_+$, the probability that it reaches $\phi_+$ before $\phi_-$, denoted $p_+(\phi_0)$ [respectively the probability that it reaches $\phi_-$ before $\phi_+$, denoted $p_-(\phi_0)$], obeys the differential equation
\bea
v p''_\pm(\phi) - \frac{v'}{v} p'_\pm(\phi) = 0\, ,
\label{original-p-eq}
\eea
with boundary conditions $p_\pm(\phi_\pm)=1$ and $p_\pm(\phi_\mp)=0$. Here the reduced potential $v$ is defined through
\bea
\label{eq:v:def}
V(\phi)=24\pi^2\Mp^4v(\phi)\, .
\eea
This equation can be solved analytically and one obtains
\bea
\label{eq:ppm:generic}
p_\pm(\phi_0) = \pm \displaystyle \dfrac{\displaystyle\int_{\phi_\mp}^{\phi_0} e^{-\frac{1}{v(\phi)}}\dd\phi}{\displaystyle\int_{\phi_-}^{\phi_+} e^{-\frac{1}{v(\phi)}}\dd\phi}\, .
\eea
One can check $p_++p_-=1$.

\begin{figure}[t]
\begin{center}
\includegraphics[width=0.45\textwidth]{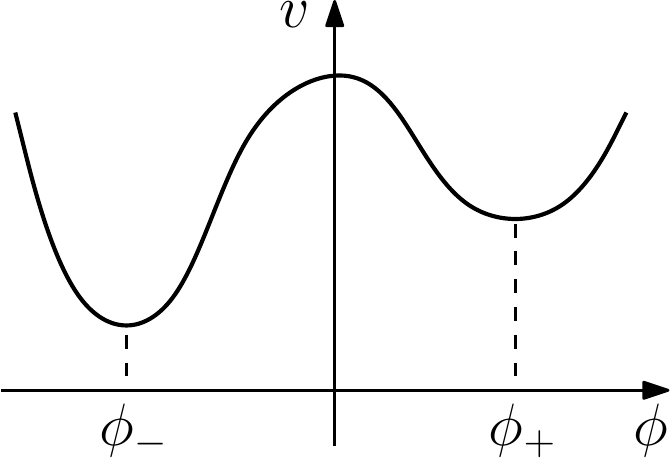}
$\quad\quad\quad$
\includegraphics[width=0.45\textwidth]{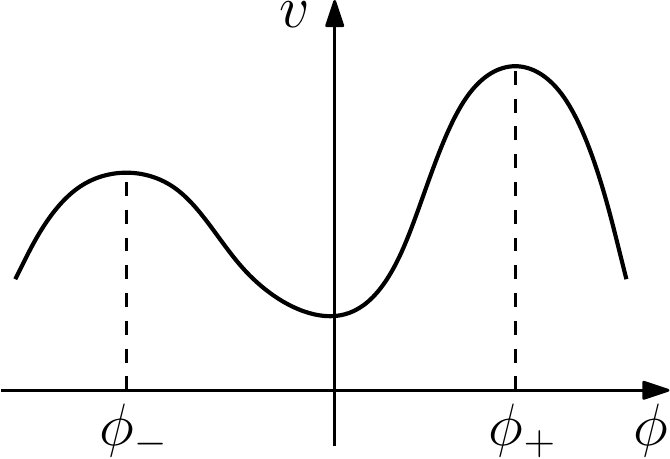}
\caption{Sketch of the potentials studied in this work, where the field falls from a local maximum towards either of two vacua (left panel, see \Sec{sec:sym-brk}), or escapes from a local minimum through either of two potential barriers (right panel, see \Sec{sec:escape}).}
\label{fig:fall-escape-pot}
\end{center}
\end{figure}
In the following we consider two types of situations depicted in \Fig{fig:fall-escape-pot}. In the first one (left panel), the inflaton field is placed at a local maximum of its potential and can fall towards either of two local minima located at $\phi_-$ and $\phi_+$. In the second configuration (right panel), the inflaton field is placed at a local minimum and escapes through either of two potential barriers, located at $\phi_-$ and $\phi_+$. Notice that in that case, we choose the boundaries $\phi_\pm$ to be located at the maxima of the potential.  In other words, we define ``escaping through the right'' as reaching the right maximum before the left one.  We could have defined it as reaching a farther point on the right (another local minimum, or even $+\infty$), but as we will show this would not substantially change our results.

For simplicity, we let $\phi_0=0$ be the initial field value of the inflaton (\ie the local maximum of the potential in the first case and the local minimum in the second), and we assume that slow roll holds in the entire range $[\phi_-,\phi_+]$. We are interested in the difference between $p_+$ and $p_-$, described by the ratio
\bea
\label{r}
R = \frac{p_+}{p_-} = \frac{\displaystyle\int_{\phi_-}^0 e^{-\frac{1}{v(\phi)}} \dd \phi}{\displaystyle\int_0^{\phi_+} e^{-\frac{1}{v(\phi)}} \dd \phi}\, .
\eea 
In particular, we want to quantify the amount of symmetry breaking in the potential that is required for having $R$ significantly different from $1$, \ie for having substantially asymmetric fall/escape probabilities.

The rest of this paper is organised as follows.  In \Sec{sec:sym-brk} we analyse the ``fall problem'' where the inflaton falls from a local maximum towards either of two minima, and in \Sec{sec:escape} we study the ``escape problem'' where the inflaton crosses barriers located around a local minimum of its potential. The time taken for these processes to happen is computed in \Sec{sec:DecayRate}. The generic case featuring multiple minima and maxima is discussed in \Sec{sec:gen-pot} and we conclude in \Sec{sec:conclusions}. In \App{app:asymp-exp}, an analytical approximation method used throughout the paper and based on the steepest descent approximation is detailed.
\section{Falling from a local maximum}
\label{sec:sym-brk}
Let us first discuss the case depicted in the left panel of \Fig{fig:fall-escape-pot}, where the field is placed at a local maximum of its potential and we calculate the probabilities that it falls towards either of the two minima located on both sides. For a given potential, it is straightforward to compute \Eq{r} numerically and an example will be discussed below. Let us first derive an analytical approximation of the result in order to discuss the different regimes that one can encounter.
\subsection{Steepest descent approximation}
\label{sec:sym-br:SharpMaxLimit}
We consider the regime where, because of the exponential form (with large arguments) of the integrands in \Eq{r}, most of the contributions to the integrals come from the neighbourhood of the maximum of the potential. There, the argument of the exponential can be Taylor expanded, and for the denominator of $R$, this leads to
\bea
\label{eq:fall:denominator}
\int_0^{\phi_+} \exp \left[ \frac{-1}{v(\phi)} \right] \dd \phi = \int_0^{\phi_+} \exp \left[ \frac{-1}{v(0)} + \frac12 \frac{v''(0)}{v(0)^2} \phi^2 + \frac1{3!} \frac{v'''(0)}{v(0)^2} \phi^3 + \ldots \right] \dd \phi\, .
\eea
In this expression, we have used that the potential is maximal at $\phi=0$, hence $v'(0)=0$ and $v''(0)<0$. The second term in the expansion, proportional to $v''(0)$, implies that most of the contribution to the integral comes from an interval of a few $\Delta\phi = v(0)/\sqrt{v''(0)}$ centred around $\phi=0$. For the Taylor series to be well behaved, the third term in the expansion, proportional to $v'''(0)$, should remain negligible until $\phi=\pm\Delta\phi$ at least, which gives rise to the condition
\bea
\label{eq:fall:cond:1}
\frac{v(0)\left\vert v'''(0)\right\vert }{\left\vert v''(0)\right\vert^{3/2}} \ll 1\, .
\eea
Then, if $\phi_+\gg \Delta\phi$, the upper bound in the integral~(\ref{eq:fall:denominator}) can be taken to infinity. This is the case if
\bea
\label{eq:fall:cond:2}
\frac{\vert v''(0) \vert}{v(0)^2} \phi_+^2\gg 1\, .
\eea
If these two conditions are satisfied, it is carefully shown in \App{app:asymp-exp} that the above integral can be approximated by
\bea
\label{eq:fall:denominator:appr}
\int_0^{\phi_+} \exp \left[ \frac{-1}{v(\phi)} \right] \dd \phi \simeq \sqrt{\frac{\pi}{2}}\frac{v(0)\ee^{-\frac{1}{v(0)}}}{\sqrt{\left\vert v''(0)\right\vert}}\left[1+\sqrt{\frac{2}{\pi}}\frac{v(0)v'''(0)}{3\left\vert v''(0)\right\vert^{3/2}}\right]\, .
\eea
A similar calculation can be performed for the numerator of \Eq{r} where only the sign in front of $v'''(0)$ changes, and $\phi_+$ has to be replaced with $\phi_-$ in \Eq{eq:fall:cond:2}. This gives rise to
\bea
\label{eq:R:fall:SharpMax:appr}
R\simeq 1 - \frac{2}{3}\sqrt{\frac{2}{\pi}}\frac{v(0)v'''(0)}{\left\vert v''(0)\right\vert^{3/2}}\, .
\eea
One notices that the condition~(\ref{eq:fall:cond:1}) for which the approximation scheme is valid precisely guarantees that $\vert R-1\vert \ll 1$.  This approximation therefore corresponds to a small symmetry breaking limit. Let us also remark that if the third derivative of the potential exactly vanishes at the top of the potential, then the leading contribution to $R-1$ comes from the first non-vanishing odd derivative of the potential function at its maximum. 
\subsection{Example}
\label{sec:fall:example}
\begin{figure}[t]
\begin{center}
\includegraphics[width=0.45\textwidth]{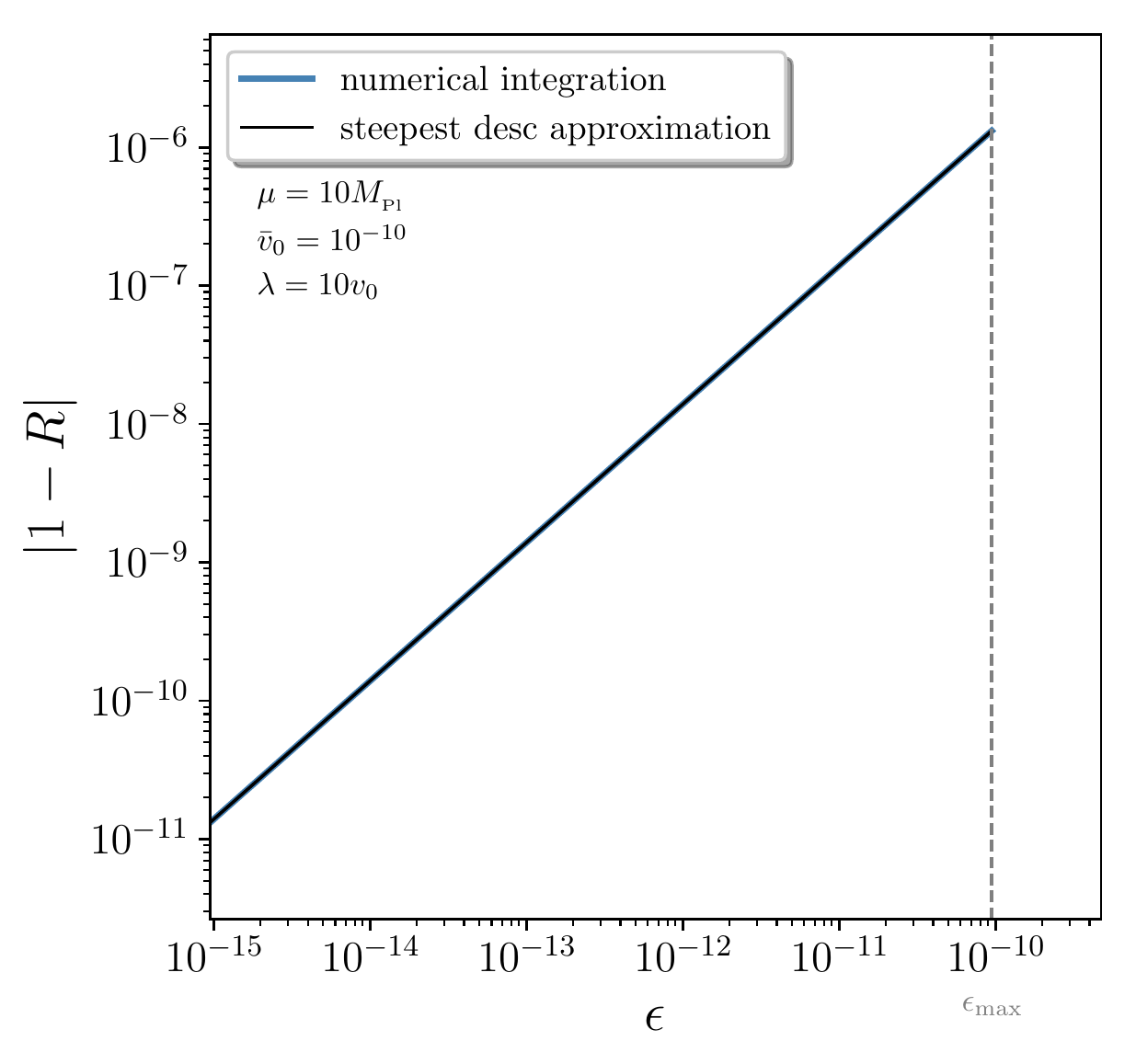}
\includegraphics[width=0.45\textwidth]{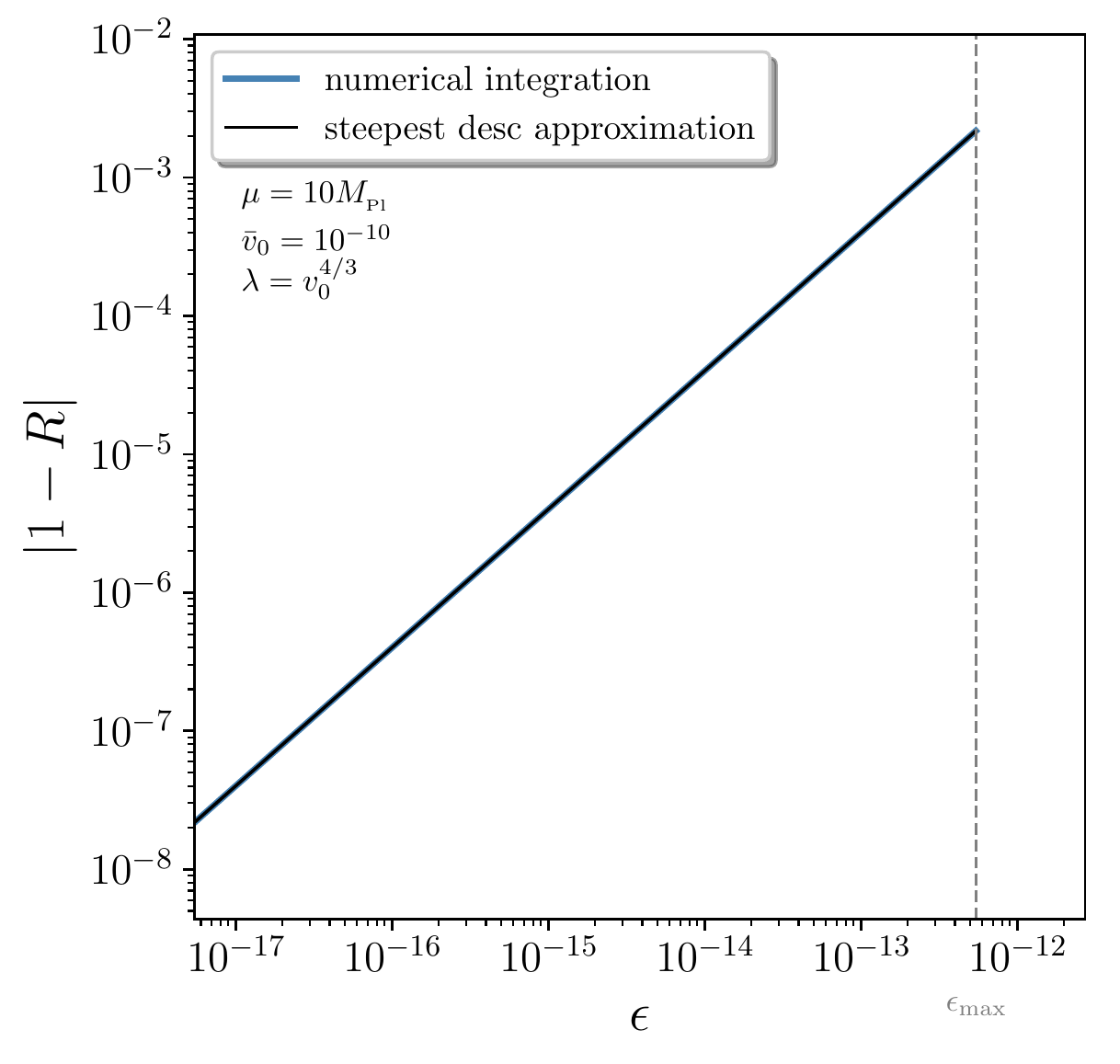}
\includegraphics[width=0.45\textwidth]{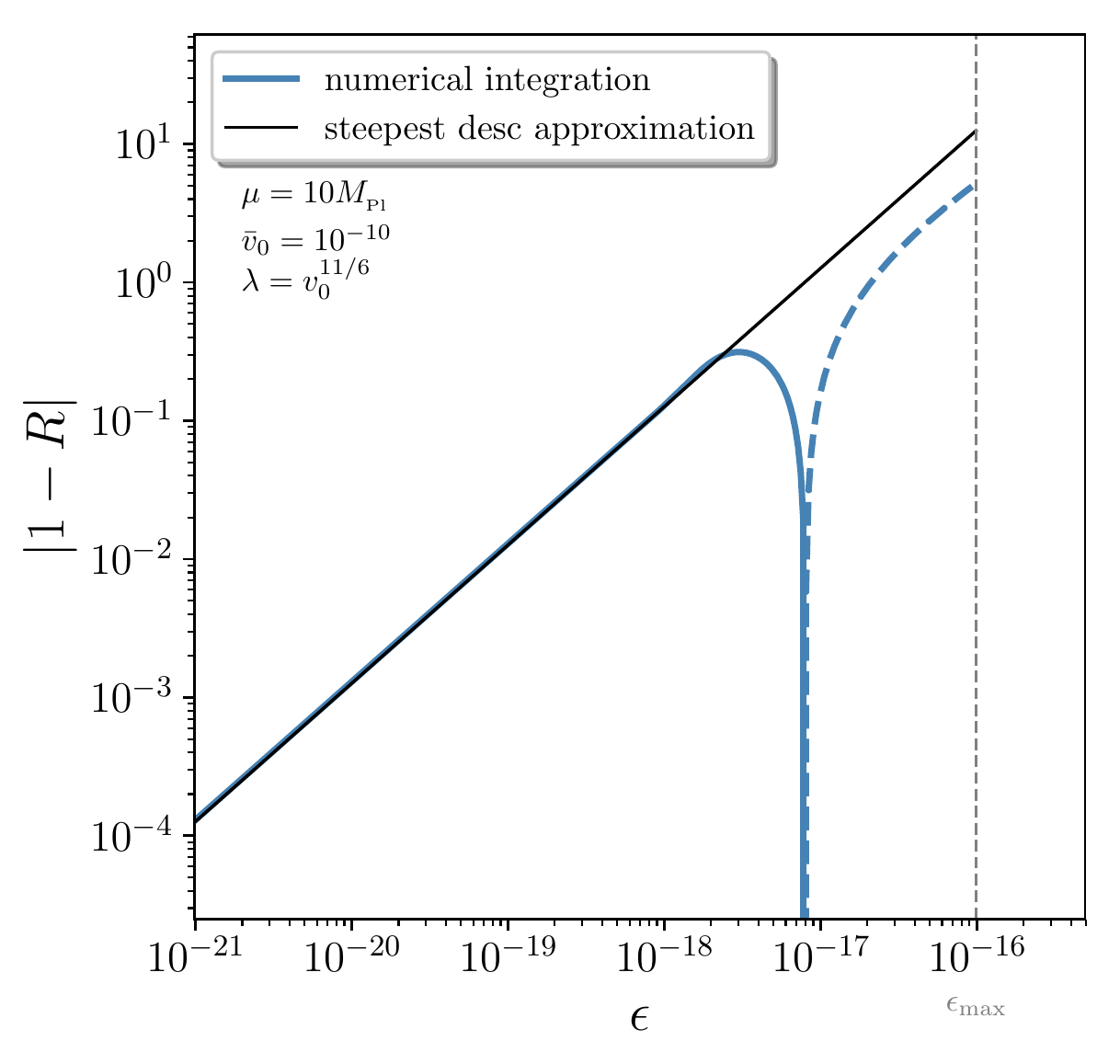}
\includegraphics[width=0.45\textwidth]{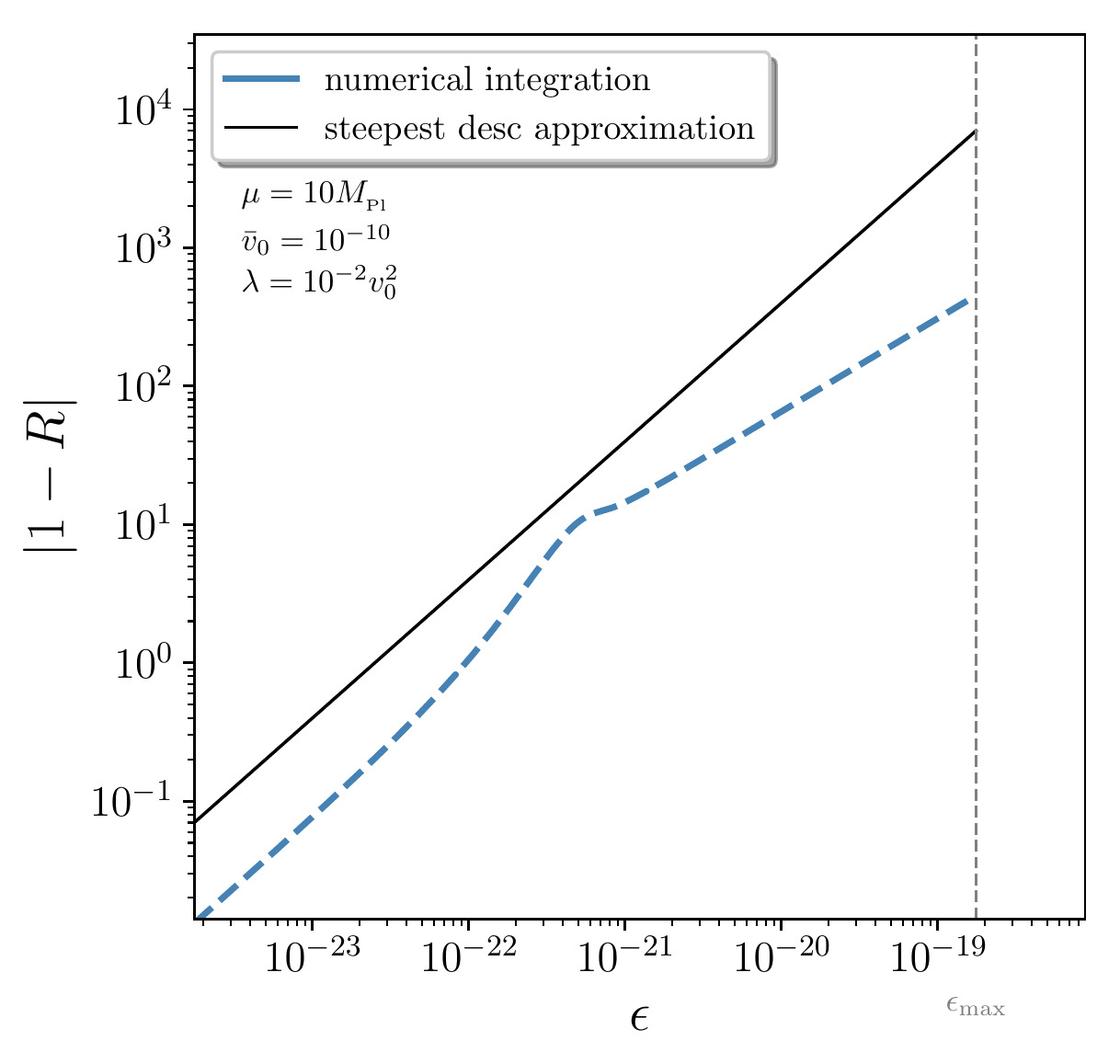}
\caption{Decay probability ratio $R=p_+/p_-$ as a function of the symmetry-breaking parameter $\epsilon$ for the model~(\ref{quartic-potential}). The blue line is computed numerically from \Eq{r} and the black line corresponds to the steepest descent approximation~(\ref{eq:BrokenMexicanHat:R:appr}). The solid part of the curves correspond to when $R<1$ while the dashed part of the curves correspond to when $R>1$. The different panels correspond to different regimes in parameter space. In the top left panel, $\lambda>\bar{v}_0$ and in the top right panel, $\bar{v}_0^{5/3}\ll \lambda \ll \bar{v}_0$. In both cases, the steepest descent approximation is always valid (the curves cannot be distinguished by eye) and $R$ is always close to one. In the bottom left panel, $\bar{v}_0^2\ll \lambda\ll \bar{v}_0^{5/3}$, and when $\epsilon$ is sufficiently large, $R$ can substantially deviate from one and the steepest descent approximation breaks down. In the bottom right panel, $\lambda\ll \bar{v}_0^2$ for which the approximation always breaks down and $R$ can substantially deviate from one if $\epsilon$ is large enough.}
\label{fig:SymBrk_r}
\end{center}
\end{figure}
Let us illustrate the above considerations with an explicit example. In general, any potential $v(\phi)$ with a local maximum at $\phi=0$ can be decomposed into an even part and an odd part around its maximum,
\bea
\label{symmetry-breaking}
v(\phi)=\bar v(\phi)+\epsilon \Delta v(\phi)\, .
\eea
In this expression, $\bar{v}(\phi)$ is an even function of $\phi$ and $\Delta v(\phi)$ is an odd function of $\phi$ with vanishing derivative at the origin (such that $\phi=0$ is a local maximum of the potential function). With such a decomposition, \Eq{eq:R:fall:SharpMax:appr} gives rise to 
\bea
\label{eq:R:fall:SharpMax:appr:EvenOdd}
R\simeq 1 -\frac{2}{3}\sqrt{\frac{2}{\pi}}\frac{\bar{v}(0)\Delta v'''(0)}{\left\vert \bar{v}''(0)\right\vert^{3/2}}  \epsilon\, .
\eea
In order to check the validity of this analytical formula, for explicitness let us consider the case where $\bar{v}$ is of the Mexican hat type with minima at $\phi=\pm\mu$ and $\Delta v$ is a cubic function,
\bea
\label{quartic-potential}
\bar{v}\left(\phi\right)=\lambda\left[ \left( \frac{\phi}{\mu} \right)^2 - 1 \right]^2 + \bar v_0,
\quad\quad\quad
\Delta v\left(\phi\right) = \left( \frac{\phi}{\mu} \right)^3\, .
\eea
The resulting potential has a local maximum at $\phi=0$ and two local minima at $\phi_{\pm}/\mu=\pm\sqrt{1+[3\epsilon/(8\lambda)]^2}-3\epsilon/(8\lambda)$. Requiring that the potential is positive at these two minima yields an upper bound on $\epsilon$ that we denote $\epsilon_{\mathrm{max}}$ and that depends on $\lambda$ and $\bar{v}_0$. An explicit expression can be derived but we do not reproduce it here since it is not particularly illuminating. Let us simply notice that, when $\bar{v}_0\ll \lambda$, $\epsilon_{\mathrm{max}}\simeq \bar{v}_0$, and when $\bar{v}_0\gg \lambda$, $\epsilon_{\mathrm{max}}\simeq 4(\lambda/3)^{3/4}v_0^{1/4} $. Slow roll also imposes that $\mu\gg \Mp$ and $\lambda/\bar{v}_0\ll \mu^2/\Mp^2$. 

The integrals appearing in \Eq{r} can be computed numerically and the result is displayed in \Fig{fig:SymBrk_r}, together with the approximation~(\ref{eq:R:fall:SharpMax:appr:EvenOdd}), which gives rise to
\bea
\label{eq:BrokenMexicanHat:R:appr}
R\simeq 1-\frac{\lambda+\bar{v}_0}{\lambda^{3/2}} \frac{\epsilon}{\sqrt{2\pi}}\, .
\eea
The validity conditions for this approximation to hold are given by \Eqs{eq:fall:cond:1} and~(\ref{eq:fall:cond:2}), namely the requirement that $\vert R-1\vert$ computed with \Eq{eq:BrokenMexicanHat:R:appr} remains small and that 
\bea
\label{eq:fall:cond:2:example}
\frac{\lambda}{\left(\lambda+\bar{v}_0\right)^2}\left(\frac{\phi_{\pm}}{\mu}\right)^2\gg 1\, .
\eea
Four different regimes need to be distinguished that correspond to the four panels in \Fig{fig:SymBrk_r}.

If  $\lambda\gg \bar{v}_0$, $\epsilon_{\mathrm{max}}\simeq \bar{v}_0$, and according to \Eq{eq:BrokenMexicanHat:R:appr}, $R-1\simeq - \epsilon/\sqrt{2\pi\lambda}$ so $\vert R-1\vert < \bar{v}_0/\sqrt{2\pi\lambda} < \sqrt{\bar{v}_0/(2\pi)}\ll 1$. Since $\epsilon_{\mathrm{max}}\simeq \bar{v}_0<\lambda$, one has $\phi_\pm\simeq\pm\mu$, and the second validity condition~(\ref{eq:fall:cond:2:example}) simply gives $\lambda\ll 1$, which is always satisfied. Therefore, $R$ always remains close to one and the steepest descent  approximation~(\ref{eq:BrokenMexicanHat:R:appr}) always provides an excellent fit to the full result, as can be checked on the top left panel of \Fig{fig:SymBrk_r}.

If $\lambda\ll \bar{v}_0$, \Eq{eq:BrokenMexicanHat:R:appr} gives rise to $R-1\simeq -\bar{v}_0\epsilon/(\lambda^{3/2}\sqrt{2\pi})$. Since $\epsilon_{\mathrm{max}}\simeq 4(\lambda/3)^{3/4}\bar{v}_0^{1/4} $ in that case, this leads to $R-1\simeq -4/\sqrt{2\pi} (3 \lambda)^{-3/4} \bar{v}_0^{5/4}  \epsilon/\epsilon_{\mathrm{max}}$, which implies that $R$ can substantially deviate from one only when $\lambda\ll \bar{v}_0^{5/3}$. 
Let us also notice that since $\epsilon_{\mathrm{max}}\gg \lambda$ in that case, $\phi_+/\mu\simeq [\lambda/(3\bar{v}_0)]^{1/4}$ when $\epsilon=\epsilon_{\mathrm{max}}$. The second validity condition~(\ref{eq:fall:cond:2:example}) thus also gives $\lambda\gg \bar{v}_0^{5/3}$. When $\epsilon=0$ on the other hand $\phi_{\pm}=\pm \mu$ and it simply gives $\lambda\gg \bar{v}_0^2$. We have therefore three possibilities. 
If $\bar{v}_0^{5/3}\ll\lambda\ll\bar{v}_0$, $R$ always remains close to one and the steepest descent approximation always works. This corresponds to the top right panel in \Fig{fig:SymBrk_r}.
If $\bar{v}_0^2\ll\lambda\ll \bar{v}_0^{5/3}$, $R$ can substantially deviate from one and the steepest descent approximation breaks down only if $\epsilon$ is large enough. This is the case displayed in the bottom left panel of \Fig{fig:SymBrk_r} where one can check that indeed, when $\epsilon$ is large enough, $R-1$ can become sizeable (and even changes sign).
Finally, if $\lambda\ll \bar{v}_0^2$, the second validity condition~(\ref{eq:fall:cond:2}) for the steepest descent approximation always breaks down. This case is shown in the bottom right panel of \Fig{fig:SymBrk_r} where one can see that $\vert R-1\vert$ can be large if $\epsilon$ is large enough, and that the steepest descent approximation does not even correctly predict the sign of $R-1$.

In summary, we find that unless $\lambda\ll \bar{v}_0^{5/3}$, no substantial asymmetry in the decay channels can be obtained in this model, and our steepest descent approximation always works. 
Let us stress that $\lambda\ll \bar{v}_0^{5/3}$ corresponds to an extremely flat potential where the relative difference between its minimal ($\sim \bar{v}_0$) and maximal ($\bar{v}_0+\lambda$) values does not exceed $\bar{v}_0^{3/2}\ll 1$.
\section{Escaping from a local minimum}
\label{sec:escape}
Let us now discuss the case depicted in the right panel of \Fig{fig:fall-escape-pot}, where the field is placed at a local minimum of its potential and escapes through one of the two potential barriers located on both sides. As in the previous section, we first make use of the steepest descent approximation to derive an analytical estimate of the ratio between the two tunnelling probabilities, before studying one numerical example in more details.
\subsection{Steepest descent approximation}
\label{sec:escape:sharp}
As in \Sec{sec:sym-br:SharpMaxLimit}, we consider the regime where, because of the exponential form of the integrands in \Eq{r} and because of their large negative arguments, most of the contributions to the integrals come from the neighbourhood of the maxima of the potential. There, the argument of the exponential can be Taylor expanded, and after changing the integration variable to $\chi=\phi_+-\phi$ in the denominator of $R$, this leads to
\bea
\label{eq:escape:denominator}
\int_0^{\phi_+} \exp \left[ \frac{-1}{v(\phi)} \right] \dd \phi = \int_0^{\phi_+} & \exp \left[ \frac{-1}{v\left(\phi_+\right)} + \frac12 \frac{v''\left(\phi_+\right)}{v\left(\phi_+\right)^2} \chi^2 
+ \frac{1}{3!}\frac{v'''\left(\phi_+\right)}{v\left(\phi_+\right)^2}\chi^3+ \ldots \right] \dd \chi\, .
\eea
In this expression, we have used the fact that since $\phi_+$ is a local maximum of the potential, $v'(\phi_+)=0$. The second term in the expansion, proportional to $v''(\phi_+)$, implies that most of the contribution to the integral comes from an interval of a few $\Delta\chi = v(\phi_+)/\sqrt{v''(\phi_+)}$ centred around $0$. For the Taylor series to be well behaved, the third term in the expansion, proportional to $v'''(0)$, should remain negligible until $\chi=\Delta\chi$ at least, which gives rise to the condition
\bea
\label{eq:escape:cond:1}
\frac{v\left(\phi_+\right)\left\vert v'''\left(\phi_+\right)\right\vert }{\left\vert v''\left(\phi_+\right)\right\vert^{3/2}} \ll 1\, .
\eea
Then, if $\phi_+\gg \Delta\phi$, the upper bound in the integral~(\ref{eq:escape:denominator}) can be taken to infinity. This is the case if
\bea
\label{eq:escape:cond:2}
\frac{\vert v''\left(\phi_+\right) \vert}{v\left(\phi_+\right)^2} \phi_+^2\gg 1\, .
\eea
If these two conditions are satisfied, it is shown in \App{app:asymp-exp} that the above integral can be approximated by
\bea
\label{eq:escape:denominator:appr}
\int_0^{\phi_+} \exp \left[ \frac{-1}{v(\phi)} \right] \dd \phi = \sqrt{\frac{\pi}{2}}\frac{v\left(\phi_+\right)\ee^{-\frac{1}{v\left(\phi_+\right)}}}{\sqrt{\left\vert v''\left(\phi_+\right)\right\vert}}\, ,
\eea
which is of course similar to \Eq{eq:fall:denominator:appr} if one evaluates the potential and its derivatives at $\phi_+$ instead of $0$, except that the leading-order result is enough here since it is already asymmetric. For the numerator of \Eq{r}, a similar calculation can be performed, where $\phi_+$ simply has to be replaced by $\phi_-$ in \Eqs{eq:escape:cond:1}-(\ref{eq:escape:denominator:appr}). For the ratio of the two tunnelling probabilities, this gives rise to
\bea
\label{eq:escape:appr}
R\simeq \sqrt{\frac{v''\left(\phi_+\right)}{v''\left(\phi_-\right)}}\frac{v\left(\phi_-\right)}{v\left(\phi_+\right)}\exp\left[\frac{1}{v\left(\phi_+\right)}-\frac{1}{v\left(\phi_-\right)}\right]\, .
\eea
Contrary to \Sec{sec:sym-brk}, one can see that the consistency conditions~(\ref{eq:escape:cond:1}) and~(\ref{eq:escape:cond:2}) do not a priori prevent $R$ from being much different from one here. Since the rescaled potential~(\ref{eq:v:def}) has to be much smaller than one, the order of magnitude of $R$ is mostly determined by the exponential term, $R\sim \ee^{1/v(\phi_+)-1/v(\phi_-)}$. This expression is consistent with the intuition that the system is more likely to escape through the shorter barrier.
\subsection{Example}
\label{subsec:escape:example}
\begin{figure}[t]
\begin{center}
\includegraphics[width=0.45\textwidth]{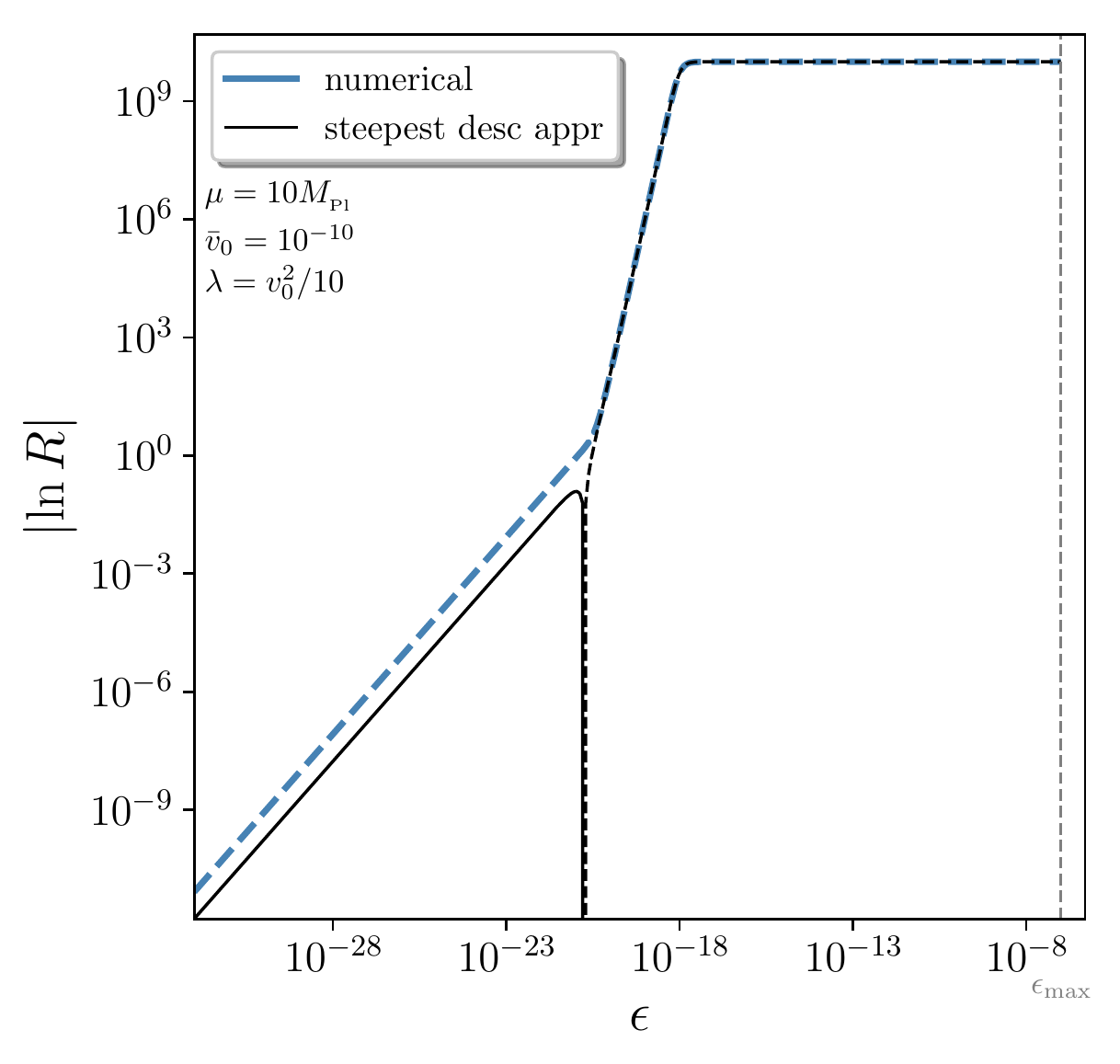}
\includegraphics[width=0.45\textwidth]{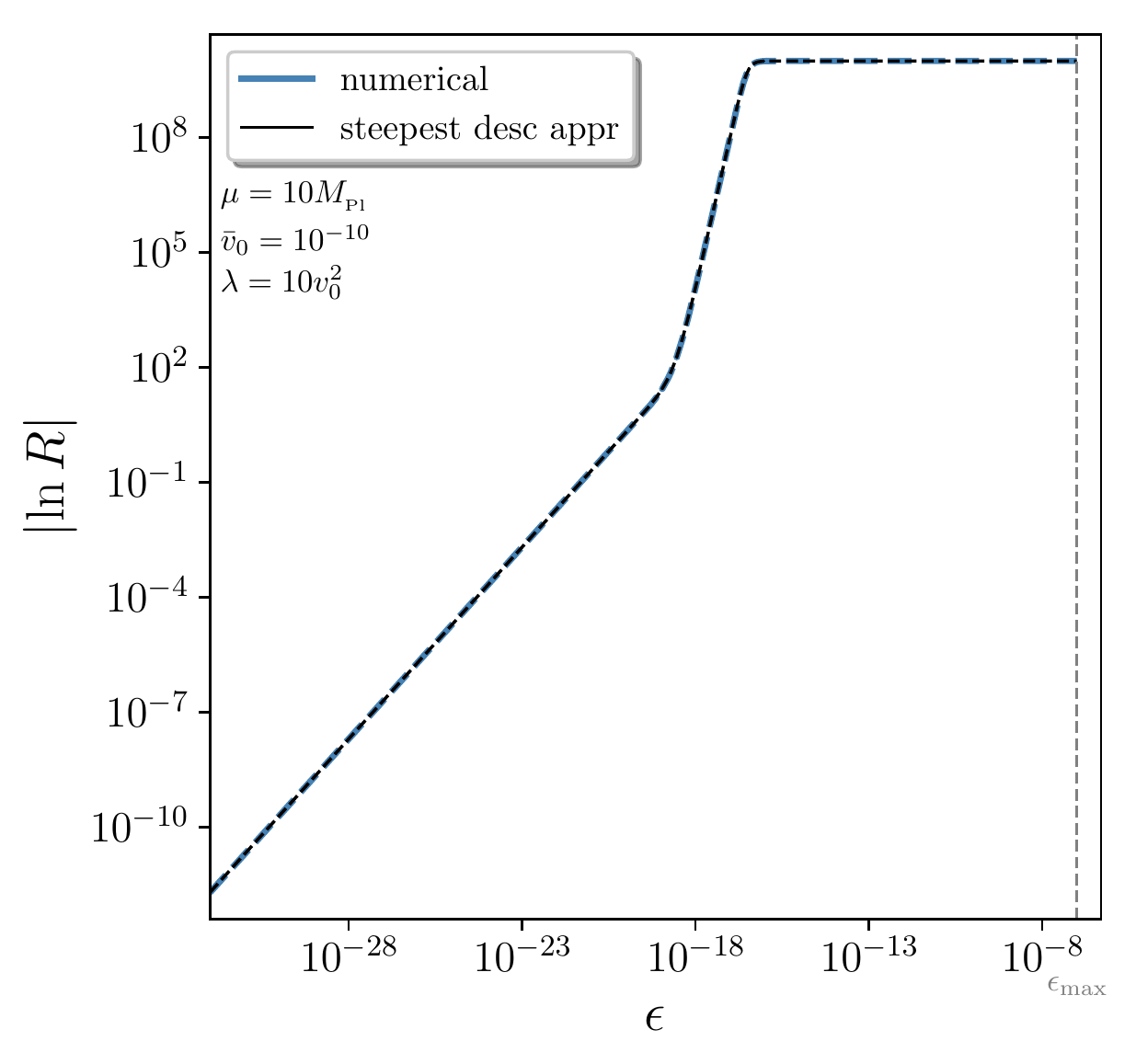}
\caption{Decay probability ratio $R=p_+/p_-$ as a function of the symmetry-breaking parameter $\epsilon$ for the model~(\ref{eq:escape:example:pot}). The blue line is computed numerically from \Eq{r} and the black line corresponds to the steepest descent approximation~(\ref{eq:R:fall:example:appr}). The solid part of the curves correspond to when $R>1$ while the dashed part of the curves correspond to when $R<1$. The different panels correspond to different regimes in parameter space. In the left panel, $\lambda<\bar{v}_0^2$ and the steepest descent approximation breaks down at $\epsilon<\bar{v}_0^2$. In the right panel, $\lambda>\bar{v}_0^2$ and the steepest descent approximation always holds. In both cases, $R$ significantly differs from one when $\epsilon>\bar{v}_0^2$. Notice that the (absolute value of the) logarithm of $R$ itself is displayed with a logarithmic scale.}
\label{fig:SymBrk_r:escape}
\end{center}
\end{figure}
In order to further illustrate the difference with the setup studied in \Sec{sec:sym-brk}, let us introduce a symmetry broken potential analogous to the one of \Sec{sec:fall:example}, made of an inverted Mexican hat even part and of a cubic odd part,
\bea
\label{eq:escape:example:pot}
v = \bar v + \epsilon \Delta v = -\lambda\left[\left(\frac{\phi}{\mu}\right)^2-1\right]^2 + \bar{v}_0 + \epsilon \left( \frac{\phi}{\mu} \right)^3\, .
\eea
This potential has a local minimum at $\phi=0$ where $v(0)=\bar{v}_0-\lambda$ and two local maxima at $\phi_{\pm}/\mu=\pm\sqrt{1+[3\epsilon/(8\lambda)]^2}+3\epsilon/(8\lambda)$. For the potential to be positive in the range $[\phi_-,\phi_+]$ under consideration, one simply has to impose $\lambda<\bar{v}_0$, otherwise there is no upper bound on $\epsilon$ apart from the slow-roll condition $\epsilon<\epsilon_{\mathrm{max}}=\left(\bar{v}_0-\lambda\right)\mu^3/\Mp^3$. Slow roll further imposes $\mu\gg \Mp$ and $\lambda/\bar{v}_0\ll \mu^2/\Mp^2$. With this potential, the steepest descent approximation, \Eq{eq:escape:appr}, gives rise to
\bea
\label{eq:R:fall:example:appr}
R\simeq & \frac{512 \bar{v}_0 \lambda^3 + 27\epsilon^4+288 \epsilon^2 \lambda^2  - \epsilon\left(9 \epsilon^2+64\lambda^2\right)^{3/2} }{512 \bar{v}_0 \lambda^3 + 27\epsilon^4+288 \epsilon^2 \lambda^2  + \epsilon\left(9 \epsilon^2+64\lambda^2\right)^{3/2}}
\sqrt{1+\frac{3\epsilon\left(3\epsilon+\sqrt{9\epsilon^2+64\lambda^2}\right)}{32\lambda^2}}
\\ & \times \exp\left[\frac{-\epsilon\left(9 \epsilon^2 + 64 \lambda^2\right)^{3/2}}{ 27 \epsilon^4\left(\bar{v}_0-\lambda\right) +288\bar{v}_0 \epsilon^2 \lambda^2 + 256 (\bar{v}_0^2 - \epsilon^2)\lambda^3}\right]\, .
\eea
In order to make this expression more explicit, it can be expanded in the small $\epsilon$ limit where one gets
\bea
\label{eq:R:fall:example:appr:appr}
\left. R \right\vert_{\epsilon\ll \lambda}\sim  \ee^{-\frac{2\epsilon}{v_0^2}}\, .
\eea
This shows that, unless the symmetry breaking parameter is tiny $\epsilon\ll v_0^2$, the tunnelling probabilities are highly asymmetric. 

The small $\epsilon$ limit can also be written for the generic even/odd decomposition of the potential given in the first equality of \Eq{eq:escape:example:pot}, since at leading order in $\epsilon$ the maximas are displaced according to $\phi_{\pm}\simeq\pm\mu\mp\epsilon\Delta v(\mu)/\bar{v}''(\mu)$. This gives $v(\phi_{\pm})\simeq \bar{v}(\mu)\pm\epsilon\Delta v(\mu) $, hence $R\sim \ee^{-2\epsilon\Delta v(\mu)/\bar{v}^2(\mu)}$, which is of course consistent with \Eq{eq:R:fall:example:appr:appr}.

The validity conditions~(\ref{eq:escape:cond:1}) and~(\ref{eq:escape:cond:2}) for the steepest descent approximation can also be verified. With the restrictions on parameters mentioned above, one can check that the first condition is always satisfied, and the second one is violated only if both $\lambda$ and $\epsilon$ are smaller than $\bar{v}_0^2$. This gives rise to two regimes displayed in the two panels of \Fig{fig:SymBrk_r:escape}. If $\lambda\ll \bar{v}_0^2$, the steepest descent approximation is only valid for $\epsilon\gg \bar{v}_0^2$. This corresponds to the left panel in \Fig{fig:SymBrk_r:escape}. If $\lambda\gg \bar{v}_0^2$, the steepest descent approximation is always valid, and this corresponds to the right panel in \Fig{fig:SymBrk_r:escape}. 

Interestingly, one notices that in the case $\lambda\ll \bar{v}_0^2$, the condition on $\epsilon$ for the steepest descent approximation to hold, $\epsilon>\bar{v}_0^2$, is precisely the one such that $R<1$ according to \Eq{eq:R:fall:example:appr:appr}. This shows that the steepest descent approximation is always valid to describe probability ratios away from one. In particular, the large $\epsilon$ limit, $\left.R\right\vert_{\epsilon\gg \lambda} \sim \ee^{-1/(\bar{v}_0-\lambda)}$, is always correctly reproduced in \Fig{fig:SymBrk_r:escape}.
%For example, suppose inflation started from $\phi_+$ and ended in $\phi=0$, which we take to be the present vacuum state.  Then we can have $\bar v_0, \lambda \sim 10^{-10}$ (their difference being $O(10^{-120})$), and take $\epsilon = 0.1 \bar v_0^2$.  (These are consistent with $\bar v_0>\lambda$ and $\bar v_0>|\epsilon|$, which are required to avoid an AdS vacuum.)  The question now is the relative probability of escaping from our current vacuum to the right or left.  Plugging in the numbers, we find 
%\begin{equation}
%R \approx 1 - \frac{2\epsilon}{\bar v_0^2} = 0.8.
%\end{equation}
%This means that we can go back from where we are ($\phi=0$) to where we came from ($\phi=\phi_+$) with probability $\frac49$, and escape to the other (lower) local maximum ($\phi=\phi_-$) with probability $\frac59$ (or vice versa). This is again a situation with significant asymmetry in the escape probabilities.
%
\section{Decay rate}
\label{sec:DecayRate}
In the two previous sections, we have calculated the relative probabilities to reach either of two potential minima starting from a local maximum, and to tunnel through either of two potential barriers from a local minimum. We now study the typical time required for this to happen. In \Ref{Vennin:2015hra}, first passage time techniques are employed to show that the mean number of \efolds~$\langle \mathcal{N} \rangle (\phi)$ required to reach either $\phi_-$ or $\phi_+$ starting from the initial field value $\phi$, obeys the differential equation
\bea
\label{eq:eom:meanN}
v\left\langle \mathcal{N}\right\rangle''\left(\phi\right)-\frac{v'}{v}\left\langle \mathcal{N}\right\rangle'\left(\phi\right)=-\frac{1}{\Mp^2}\, ,
\eea
with boundary conditions $\langle \mathcal{N} \rangle (\phi_-)=\langle \mathcal{N} \rangle (\phi_+)=0$. This equation has the same structure as \Eq{original-p-eq}, the only difference being the non-vanishing right-hand side. It can be solved as~\cite{Vennin:2015hra}
\bea
\label{eq:meanN:generic:implicitConstant}
\left\langle \mathcal{N}\right\rangle\left(\phi\right)=\int_{\phi_-}^\phi\frac{\dd x}{\Mp}\int_{x}^{\bar{\phi}\left(\phi_-,\phi_+\right)}\frac{\dd y}{\Mp}\frac{1}{v\left(y\right)}\exp\left[\frac{1}{v\left(y\right)}-\frac{1}{v\left(x\right)}\right]\, ,
\eea
where $\bar{\phi}(\phi_-,\phi_+)$ is an integration constant that is implicitly set through the boundary condition $\langle \mathcal{N} \rangle (\phi_+)=0$. Alternatively, an explicit formula that does not rely on implicit integration constants is given by~\cite{Vennin:2015hra}
\bea
\label{eq:meanN:generic}
\left\langle \mathcal{N}\right\rangle\left(\phi\right)=\int_{\phi_-}^{\phi_+}\frac{\dd x}{\Mp}\int_{x}^{\phi_+}\frac{\dd y}{\Mp}\frac{1}{v\left(x\right)}\exp\left[\frac{1}{v\left(x\right)}-\frac{1}{v\left(y\right)}\right]\left[\theta\left(y-\phi\right)-p_-\left(\phi\right)\right]\, ,
\eea
where $\theta(y-\phi)=1$ if $y>\phi$ and $0$ otherwise, and the function $p_-(\phi)$ has been given in \Eq{eq:ppm:generic}. One can check that both \Eqs{eq:meanN:generic:implicitConstant} and~(\ref{eq:meanN:generic}) satisfy \Eq{eq:eom:meanN} and its boundary conditions.
\subsection{Escaping from a local minimum}
Let us investigate the case depicted in the right panel of \Fig{fig:fall-escape-pot} and studied in \Sec{sec:escape} first, since it will allow us to discuss the case of the left panel of \Fig{fig:fall-escape-pot} afterwards. We again make use of the steepest descent approximation, starting from \Eq{eq:meanN:generic} that we evaluate at $\phi=0$. For the integral over $y$, over the integration range the potential has a local maximum at $\phi_+$ and another one at $x$ only if $x<0$. The dominant contributions to the integral over $y$ are thus given by
\bea
\int_{x}^{\phi_+}\dd y\ee^{-\frac{1}{v\left(y\right)}}\left[\theta\left(y\right)-p_-\right]\simeq&
\left(1-p_-\right)\int_{\phi_+-\cdots}^{\phi_+}\dd y\ee^{-\frac{1}{v\left(y\right)}}
%\\ &
-\theta(-x) p_-\int_{x}^{x+\cdots}\dd y\ee^{-\frac{1}{v\left(y\right)}}\, .
\eea
In this expression, $\int_{\phi_+-\cdots}^{\phi_+}$ means that the integration is performed in the left neighbourhood of $\phi_+$, and similarly for $\int_{x}^{x+\cdots}$. Since $\phi_+$ is a local maximum of the potential, the first integral can be carried out using the techniques explained in \App{app:asymp-exp}. The second integral can be performed in a similar manner, the only difference being that since $x$ is not a local maximum of the potential in general, terms involving $v'(x)$ also have to be taken into account and one obtains
\bea
\label{eq:decayrate:Iy}
&\int_{x}^{\phi_+}\dd y\ee^{-\frac{1}{v\left(y\right)}}\left[\theta\left(y\right)-p_-\right]\simeq 
\left(1-p_-\right)
\sqrt{\frac{\pi}{2}}\frac{v\left(\phi_+\right)}{\sqrt{\left\vert v''\left(\phi_+\right)\right\vert}}\ee^{-\frac{1}{v\left(\phi_+\right)}}
\\ & \quad\quad\quad\quad\quad
-\theta(-x) p_-
\sqrt{\frac{\pi}{2}}\frac{v\left(x\right)}{\sqrt{\left\vert v''\left(x\right)\right\vert}}\ee^{-\frac{1}{v\left(x\right)}}
\frac{f\left[\frac{-v'(x)/v^2(x)}{\sqrt{2v'^2(x)/v^3(x)-2v''(x)/v^2(x)}}\right]}{\sqrt{1-\frac{v'^2(x)}{v(x)v''(x)}}}
\, ,
\eea
where $f(z)\equiv \erfc(z)\ee^{z^2}$ is such that $f(z)<1$ for $z>0$, $\erfc$ being the complementary error function. The integral over $x$ can then be performed. Since the first term in \Eq{eq:decayrate:Iy} does not depend on $x$, when integrated over $x$ it yields $\int_{\phi_-}^{\phi_+}\dd x \ee^{1/v(x)}/v(x)\dd x$. Because of the different sign in the exponential, this time the integrand is maximal when the potential is minimal, \ie around $x=0$. Still, the same steepest descent approximation as previously employed can be used and one obtains $\int_{\phi_-}^{\phi_+}\dd x \ee^{1/v(x)}/v(x)\dd x \simeq \sqrt{2\pi/v''(0)}\ee^{1/v(0)}$. For the second term, once multiplied by $\ee^{1/v(x)}/v(x)$ it gives a contribution to the integrand of order $1/\sqrt{\vert v'' \vert}$, which is negligible compared to the one from the first term. It can therefore be neglected, and recalling that $1-p_-=p_+$, one obtains
\bea
\left\langle \mathcal{N}\right\rangle\simeq
\frac{\pi v\left(\phi_+\right) p_+}{\Mp^2 \sqrt{v''\left(0\right)\left\vert v''\left(\phi_+\right)\right\vert}}\ee^{\frac{1}{v\left(0\right)}-\frac{1}{v\left(\phi_+\right)}}\, .
\eea

The probability $p_+$ can also be approximated with the steepest descent approximation, as was already done in \Sec{sec:escape:sharp}, see \Eq{eq:escape:appr} where $p_+$ is given in terms of $R$ simply by $p_+=R/(1+R)$. This finally gives rise to
\bea
\label{eq:Ndecay:escape:appr:final}
\left\langle \mathcal{N}\right\rangle \simeq \dfrac{\pi \ee^{\frac{1}{v(0)}}}{\dfrac{\Mp^2\sqrt{v''(0)\left\vert v''\left(\phi_+\right)\right\vert}}{v\left(\phi_+\right)}\ee^\frac{1}{v\left(\phi_+\right)}+\dfrac{\Mp^2\sqrt{v''(0)\left\vert v''\left(\phi_-\right)\right\vert}}{v\left(\phi_-\right)}\ee^\frac{1}{v\left(\phi_-\right)}}\, .
\eea
This expression is invariant under exchanging $\phi_+$ and $\phi_-$, which is consistent. One can also check that in the case of a symmetric potential, one obtains $\langle \mathcal{N} \rangle = \pi v(\phi_+)\ee^{1/v(0)-1/v(\phi_+)}/(2\Mp^2\sqrt{v''(0)\vert v''(\phi_+)\vert})$. The same expression can be derived by performing the steepest descent approximation in \Eq{eq:meanN:generic:implicitConstant}, where the integration constant simply reads $\bar{\phi}=0$ for a symmetric potential. Since the rescaled potential has to be much smaller than one, the order of magnitude of $\langle \mathcal{N} \rangle$ is set by the exponential term $\langle \mathcal{N} \rangle \sim \ee^{1/v(0)-1/v(\phi_+)}$, and can be very large as soon as the relative height of the potential barrier, $\Delta v/v(0)$, is not smaller than the rescaled potential itself $v(0)$. 

In the opposite case of a strongly asymmetric potential, say $v(\phi_-)\gg v(\phi_+)$, \Eq{eq:Ndecay:escape:appr:final} boils down to $\langle \mathcal{N} \rangle = \pi v(\phi_+)\ee^{1/v(0)-1/v(\phi_+)}/(\Mp^2\sqrt{v''(0)\vert v''(\phi_+)\vert})$, which is exactly two times the expression obtained for a symmetric potential. In this limit indeed, all stochastic realisations escape through the shorter potential barrier, and the decay time is simply doubled. 

This expression gives rise to a decay rate, $\Gamma_{\phi_0\rightarrow\phi_{\pm}}\sim 1/\langle  \mathcal{N}  \rangle \propto \ee^{1/v(\phi_\pm)-1/v(\phi_0)}$, that is consistent with the one obtained by Hawking and Moss~\cite{Hawking:1981fz}\footnote{More precisely, \Ref{Hawking:1981fz} computes the decay rate in terms of cosmic time while we use the number of \efolds~as the time variable. However, our analysis can be reproduced with cosmic time, starting from the Langevin equation
\bea
\label{eq:Langevin:t}
\frac{\dd\tilde{\phi}}{\dd t} = -\frac{V'}{3H}+\frac{H^{3/2}}{2\pi}\xi\left(t\right)\, ,
\eea
where $\xi(t)$ is a white Gaussian noise that is now normalised with respect to $t$, \ie $\langle \xi(t) \xi(t') \rangle = \delta(t-t')$. Here we use the notation $\tilde{\phi}$ to stress the fact that $\tilde{\phi}$ in \Eq{eq:Langevin:t} does not describe the same stochastic process as $\phi$ in \Eq{eq:Langevin}. For the boundary crossing probabilities, \Eq{eq:Langevin:t} gives rise to the same equation~(\ref{original-p-eq}) as the one obtained from \Eq{eq:Langevin}, hence the results of \Secs{sec:sym-brk} and~\ref{sec:escape} with the number of \efolds~still apply if cosmic time is used instead. For the decay rate however, \Eq{eq:eom:meanN} becomes for the mean cosmic time $\mathfrak{t}$
\bea
\label{eq:eom:meanN:t}
v^{3/2}\left\langle \mathfrak{t}\right\rangle''\left(\phi\right)-\frac{v'}{\sqrt{v}}\left\langle  \mathfrak{t}\right\rangle'\left(\phi\right)=-\frac{1}{2^{3/2}\pi\Mp^2}\, ,
\eea
which gives rise to solutions similar to \Eqs{eq:meanN:generic:implicitConstant} and~(\ref{eq:meanN:generic}), namely
\bea
\left\langle \mathfrak{t}\right\rangle\left(\phi\right)&=\frac{1}{2^{3/2}\pi\Mp}\int_{\phi_-}^\phi\frac{\dd x}{\Mp}\int_{x}^{\bar{\phi}\left(\phi_-,\phi_+\right)}\frac{\dd y}{\Mp}\frac{1}{v^{3/2}\left(y\right)}\exp\left[\frac{1}{v\left(y\right)}-\frac{1}{v\left(x\right)}\right]\\
&=\frac{1}{2^{3/2}\pi\Mp}\int_{\phi_-}^{\phi_+}\frac{\dd x}{\Mp}\int_{x}^{\phi_+}\frac{\dd y}{\Mp}\frac{1}{v^{3/2}\left(x\right)}\exp\left[\frac{1}{v\left(x\right)}-\frac{1}{v\left(y\right)}\right]\left[\theta\left(y-\phi\right)-p_-\left(\phi\right)\right]\, .
\eea
Performing the steepest descent approximation in these expressions, one finds, instead of \Eq{eq:Ndecay:escape:appr:final},
\bea
\label{eq:tdecay:escape:appr:final}
\left\langle \mathfrak{t}\right\rangle \simeq
\frac{1}{\Mp^3 2^{3/2}\sqrt{v''(0)v(0)}}
 \dfrac{\ee^{\frac{1}{v(0)}}}{\dfrac{\sqrt{\left\vert v''\left(\phi_+\right)\right\vert}}{v\left(\phi_+\right)}\ee^\frac{1}{v\left(\phi_+\right)}+\dfrac{\sqrt{\left\vert v''\left(\phi_-\right)\right\vert}}{v\left(\phi_-\right)}\ee^\frac{1}{v\left(\phi_-\right)}}\, .
\eea
This gives rise to the decay rate $\tilde{\Gamma}_{\phi_0\rightarrow\phi_{\pm}}\simeq 2^{3/2}\Mp^3\sqrt{v''(0)\vert v''(\phi_\pm)\vert v(0)}\ee^{1/v(\phi_\pm)-1/v(\phi_0)}$ that can now be directly compared with the one found in \Ref{Hawking:1981fz}.  See also Refs.~\cite{Linde:1991sk,Rey:1986zk} for different derivations.
\label{footnote:Nversust} }. 
Let us also notice that, with the above expressions, the ratio of the two decay rates, $\Gamma_{\phi_0\rightarrow\phi_{+}}/\Gamma_{\phi_0\rightarrow\phi_{-}} $ (that is independent of whether $N$ or $t$ is used as the time variable, see footnote~\ref{footnote:Nversust}), is exactly identical to the decay probability ratio $R=p_+/p_-$ derived in \Eq{eq:escape:appr} (that is also independent of the choice of time variable, see footnote~\ref{footnote:Nversust}). This ratio therefore provides a generic description of the asymmetry between the two decay channels. 

It is notable that the Hawking-Moss decay rate depends on the values of the potential and its derivatives only at the end points, and not on the detailed shape of the potential in between.  Our approach sheds some light on this:  In general, all quantities like $R$, $p_\pm$ and $\langle{\cal N}\rangle$ depend on the shape of the potential; see for example Eq.~\eqref{r}.  However, due to the exponential sensitivity of these quantities to the shape of the potential in the neighbourhood of the extrema, under the steepest descent approximation conditions~\eqref{eq:escape:cond:1} and \eqref{eq:escape:cond:2}, the dominant contribution to the relevant integrals comes from these neighbourhoods.  If the steepest descent approximation is violated, as explained in the example of Subsection~\ref{subsec:escape:example} (see the left panel of \Fig{fig:SymBrk_r:escape}), then the Hawking-Moss formula is no longer valid and the general formula \eqref{r} should be used which depends on the detailed shape of the potential.

Let us finally mention that Coleman and de~Luccia have also studied vacuum decay events \cite{Coleman:1980aw}, although in a different context. In our work, the potential barrier crossing takes place continuously under the influence of a classical noise that models the entry of super-Hubble scales into the coarse-grained sector. Equivalently, the coarse-grained field can be thought of as a classical field in the thermal bath of the de-Sitter space, which is characterised by the Hawking temperature $H/2\pi$.  In this picture, everything is ``classical'', but thermal (even if the origins of the thermal fluctuations is quantum mechanical). In particular, the barrier penetration is done by successive small jumps that get the field past the barrier ``from above''. In contrast, \Ref{Coleman:1980aw} deals with a tunnelling event ``under the barrier'' that is similar to usual tunnelling in quantum mechanics and that cannot be modelled classically. This is because the bubble size in that case is sub-Hubble (while ours is super-Hubble), and direct quantum tunnelling is more efficient than the one driven by thermal fluctuations in that regime. Furthermore, the result of \Ref{Coleman:1980aw} is applicable in the thin-wall limit, far from slow roll where we derived ours.
\subsection{Falling from a local maximum}
\label{sec:Ndecay:fall}
In the case displayed in the left panel of \Fig{fig:fall-escape-pot} where the field falls down from a local maximum to either of two minima located on both sides, a steepest descent approximation of \Eq{eq:meanN:generic} can still be performed, and one finds~\cite{Vennin:2015hra} that the mean decay time into each vacuum is basically given by the usual classical slow-roll formula $\langle \mathcal{N}\rangle_\mathrm{fall} = \int_{\phi_{\pm}}^0 \dd x v(x)/[\Mp^2 v'(x)]$.

Once the system has reached either of the two minima, the average time it takes to cross the potential barrier between the two is given by \Eq{eq:Ndecay:escape:appr:final} if one takes one maximum at $\phi=0$ and the other one at infinity, \ie
\bea
\label{eq:Ncross:fall}
\left\langle \mathcal{N}\right\rangle_{\phi_\pm\rightarrow\phi_\mp} \simeq \dfrac{\pi v(0) \ee^{\frac{1}{v\left(\phi_\pm\right)}-\frac{1}{v(0)}}}{\Mp^2\sqrt{v''(\phi_\pm)\left\vert v''\left(0\right)\right\vert}}\, .
\eea
Because of the exponential factor, this number is typically very large.
\subsection{Comparison with equilibrium distribution}
The stochastic process described by the Langevin equation~(\ref{eq:Langevin}) gives rise to a Fokker-Planck equation for the probability density $P(\phi,N)$ to find the field at value $\phi$ at time $N$, that reads~\cite{Starobinsky:1986fx} $\partial P/\partial N = \Mp^2\partial/\partial\phi (v' P/v) +\Mp^2\partial^2/\partial\phi^2(v P)$. This equation admits a stationary solution, $\partial P_{\mathrm{stat}}/\partial N=0$, given by~\cite{Starobinsky:1986fx}
\bea
\label{eq:Pstat}
P_{\mathrm{stat}}\left(\phi\right)\propto\frac{1}{v\left(\phi\right)}\exp\left[\frac{1}{v\left(\phi\right)}\right]\, .
\eea
According to that distribution, the ratio between the probability $p_{+,\mathrm{stat}}$ to lie in the vacuum located at $\phi_+$ at late time and the probability $p_{-,\mathrm{stat}}$ to lie in the vacuum located at $\phi_-$ is given by
\bea
\label{eq:Rstat}
R_{\mathrm{stat}} = \dfrac{\displaystyle\int_{0}^\infty \frac{1}{v\left(\phi\right)}\exp\left[\frac{1}{v\left(\phi\right)}\right] \dd \phi }{\displaystyle\int^{0}_{-\infty} \frac{1}{v\left(\phi\right)}\exp\left[\frac{1}{v\left(\phi\right)} \right]\dd \phi}\, .
\eea
A steepest descent expansion of this formula can be performed, and using the techniques detailed in \App{app:asymp-exp}, one obtains
\bea
\label{eq:Rstat:appr}
R_{\mathrm{stat}} \simeq \sqrt{\frac{v''\left(\phi_-\right)}{v''\left(\phi_+\right)}}\exp\left[\frac{1}{v\left(\phi_+\right)}-\frac{1}{v\left(\phi_-\right)}\right]\, .
\eea
Two remarks are in order regarding this formula.

First, it is consistent with \Eq{eq:Ncross:fall} in the following sense. In the stationary state, the fraction of the stochastic processes that lie in the vacuum centred around $\phi_-$ is given by $p_{-,\mathrm{stat}}$ and the decay rate towards the vacuum centred around $\phi_+$ is inversely proportional to $\langle \mathcal{N}\rangle_{\phi_-\rightarrow\phi_+}$, so the flux of processes that cross the potential barrier rightwards is proportional to $p_{-,\mathrm{stat}}/\langle \mathcal{N}\rangle_{\phi_-\rightarrow\phi_+}$. Similarly, the flux of processes that cross the potential barrier leftwards is proportional to $p_{+,\mathrm{stat}}/\langle \mathcal{N}\rangle_{\phi_+\rightarrow\phi_-}$. Since the distribution is stationary, the two fluxes must exactly compensate each other. This leads to $R_{\mathrm{stat}}=p_{+,\mathrm{stat}}/p_{-,\mathrm{stat}}=\langle \mathcal{N}\rangle_{\phi_+\rightarrow\phi_-}/\langle \mathcal{N}\rangle_{\phi_-\rightarrow\phi_+}$. Plugging \Eq{eq:Ncross:fall} into this formula, one exactly recovers \Eq{eq:Rstat:appr}.

Second, \Eq{eq:Rstat} is a priori very much different from \Eq{r}, and by comparing their steepest descent approximated versions, \Eqs{eq:Rstat:appr} and~(\ref{eq:R:fall:SharpMax:appr}), one notices that indeed, $R$ is typically very close to one while $R_\mathrm{stat}$ is typically very different from one because of the exponential term. The question then is: at the end of inflation, which probability ratio correctly describes the fraction of space that lies in each vacuum? This depends on the total duration of inflation. After one fall-down time $\langle \mathcal{N}\rangle_\mathrm{fall} $, given at the beginning of \Sec{sec:Ndecay:fall}, the ratio between the two vacua  populations is given by $R$ in \Eq{eq:R:fall:SharpMax:appr}. Then, the stochastic processes can go from one vacuum the other, and when $N\gg \langle \mathcal{N}\rangle_{\phi_\pm\rightarrow\phi_\mp}$ given by \Eq{eq:Ncross:fall}, this ratio converges to the stationary value given by $R_\mathrm{stat}$ in \Eq{eq:Rstat:appr}. However, the equilibration time scale~(\ref{eq:Ncross:fall}) towards the stationary distribution~(\ref{eq:Pstat}) is typically very large (at observable scales, $v\lesssim 10^{-10}$ so $\langle \mathcal{N}\rangle_{\phi_\pm\rightarrow\phi_\mp} \sim \ee^{10^{10}}$). Unless inflation lasts for a gigantic number of \efolds, the relative vacua populations at the end of inflation is therefore given by $R$, not $R_{\mathrm{stat}}$.
\section{Tunnelling in a generic potential}
\label{sec:gen-pot}
\begin{figure}
\begin{center}
\includegraphics[width=0.70\textwidth]{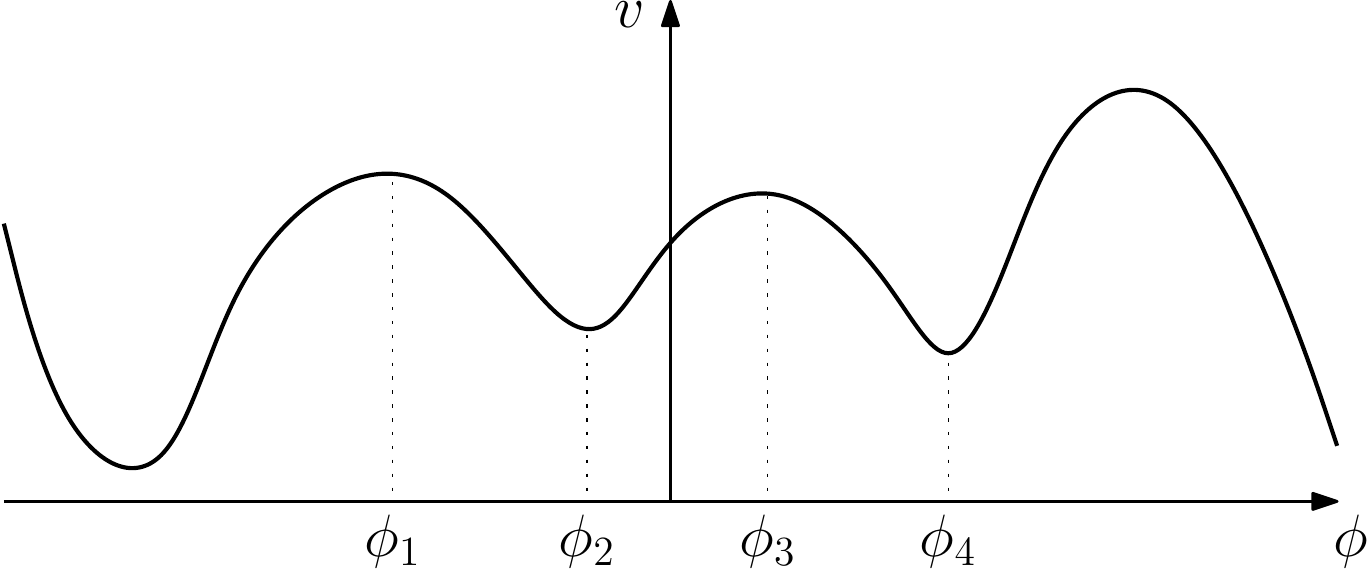}
\caption{Sketch of the generic potential discussed in \Sec{sec:gen-pot}.}
\label{fig:generic-pot}
\end{center}
\end{figure}
Let us now consider the case of a generic potential with several minima and maxima as the one sketched in \Fig{fig:generic-pot}. We will show that the results derived in the previous sections are building blocks that can be readily assembled to tackle more complex situations such as this one.

Let us first denote by $p_{ijk}$ the probability of reaching $\phi_i$ before reaching $\phi_k$, starting from $\phi_j$.  For example, the quantity $p_+$ studied in the previous sections is $p_{\phi_+0\phi_-}$, and $p_-=p_{\phi_-0\phi_+}$. The formula~(\ref{eq:ppm:generic}) can be recast as 
\bea
\label{eq:pijk}
p_{ijk} = \frac{I_j^k}{I_i^k}\, ,
\eea
where $I_i^j$ is a shorthand notation for $\int_{\phi_i}^{\phi_j}e^{-1/v}d\phi$. 

Because of the integral structure of the result, which implies that $I_j^k=I_j^\ell+I_\ell^k$, the probabilities $p_{ijk}$ obtained from \Eq{eq:pijk} obey the axiom of conditional probability, \ie they provide us with a well-defined set of probabilities that can be combined with the standard algebraic rules of probability theories. To illustrate this statement, let us consider $p_{124}$ in the potential sketched in \Fig{fig:generic-pot}.  There are two possibilities to reach $\phi_1$ before $\phi_4$ and starting from $\phi_2$: the first one is to go directly from $\phi_2$ to $\phi_1$ without passing through $\phi_3$ (this has probability $p_{123}$), and the second is to reach $\phi_3$  (this has probability $p_{321}$) and then to reach $\phi_1$ before visiting $\phi_4$ (this has probability $p_{134}$).  Therefore, one should have
\bea
\label{p124}
p_{124} = p_{123} + p_{321} p_{134}\, .
\eea
One can check that this is indeed the case by making use of \Eq{eq:pijk} and the relation $I_j^k=I_j^\ell+I_\ell^k$, which give rise to
\bea
p_{123} + p_{321} p_{134} & = \frac{I_2^3}{I_1^3} + \frac{I_2^1}{I_3^1} \frac{I_3^4}{I_1^4} 
= \frac{I_2^3 I_1^4 - I_2^1 I_3^4}{I_1^3 I_1^4}
= \frac{I_2^3 \left(I_1^2+I_2^4\right) - I_2^1 \left(I_3^2 + I_2^4\right)}{I_1^3 I_1^4}
\\
& = \frac{I_2^4\left(I_2^3 - I_2^1\right)}{I_1^3 I_1^4}
=  \frac{I_2^4}{I_1^4} = p_{124}\, .
\eea
This result obviously remains true if one replaces $1$, $2$, $3$ and $4$ with arbitrary indices $i$, $j$, $k$ and $\ell$.

The fact that the tunnelling probabilities can be manipulated according to the standard algebraic rules of probability theories imply that any tunnelling probability can be obtained from the building blocks of \Sec{sec:sym-brk} and~\ref{sec:escape}. To illustrate this second statement, let us consider the example of $p_{134}$ in \Fig{fig:generic-pot}.  From \Eq{eq:pijk} and the relation $I_j^k=I_j^\ell+I_\ell^k$ it is easy to show that $p_{ijk}+p_{kji}=1$, so $p_{134} + p_{431} = 1$ and $p_{124} + p_{421} = 1$. Then, relabelling the indices in \Eq{p124}, one can show that $p_{431} = p_{432} + p_{234} p_{421}$. Using these three equations together with \Eq{p124}, one obtains a system of four equations for the four probabilities $p_{124}$, $p_{421}$, $p_{134}$, and $p_{421}$, which can be solved and one finds
\bea
p_{134} = \frac{p_{123} p_{234}}{1 - p_{321}p_{234}}\, .
\eea
This formula expresses $p_{134}$ entirely in terms of ``fall'' ($p_{234}$) and ``escape'' ($p_{123}$ and $p_{321}$) probabilities of the kind computed in \Secs{sec:sym-brk} and~\ref{sec:escape} respectively. Let us also note that the above result can be obtained from the integral structure of \Eq{eq:pijk} directly, since
\bea
\frac{p_{123} p_{234}}{1 - p_{321}p_{234}}=
\frac{\frac{I_2^3}{I_1^3} \frac{I_3^4}{I_2^4}}{1 - \frac{I_1^2}{I_1^3} \frac{I_3^4}{I_2^4}} = \frac{I_2^3 I_3^4}{(I_1^2+I_2^3)(I_2^3+I_3^4) - I_1^2 I_3^4} = \frac{I_2^3 I_3^4}{I_2^3(I_1^2+I_2^3+I_3^4)} = \frac{I_3^4}{I_1^4}=p_{134}\, .
\eea
\section{Conclusions}
\label{sec:conclusions}
Let us now summarise our main results. We have studied tunnelling probabilities in single-field slow-roll inflationary potentials, making use of the stochastic inflation formalism~\cite{Starobinsky:1986fx} and of the first passage time techniques of \Refs{Vennin:2015hra, Assadullahi:2016gkk, Vennin:2016wnk}. We have analysed two special cases of interest, one where the inflaton falls down from a local maximum towards either of two local minima (left panel of \Fig{fig:fall-escape-pot}), and one where the inflaton escapes from a local minimum through either of two potential barriers (right panel of \Fig{fig:fall-escape-pot}). 

For the ``fall'' problem, we have found that, in the regime of validity of the steepest descent approximation scheme developed in \App{app:asymp-exp}, the asymmetry in the probabilities to decay towards each vacuum is minuscule. We have confirmed the validity of the analytical approximation~(\ref{eq:R:fall:SharpMax:appr}) that one can derive in this regime by numerically investigating the example of a double-well quartic potential perturbed by a cubic symmetry breaking term. We have found that only if the relative height $\Delta v/v$ of the unbroken potential is smaller than $v^{3/2}$ (a tiny number as soon as the process takes place at sub-Planckian energies) can the approximation be violated and the asymmetry be non-negligible.

For the ``escape problem'', since the probability to tunnel through a potential barrier depends exponentially on its potential height, we have found that the asymmetry in the tunnelling probabilities are typically much larger. In the regime of validity of the steepest descent approximation, the analytical formula~(\ref{eq:escape:appr}) was derived, and verified on a similar example as for the ``fall problem''. This allowed us to confirm again our analytical result and to show that the approximation breaks down only if the relative height $\Delta v/v$ of the potential barriers is smaller than $v^{3/2}$ and if the asymmetry is tiny.

We have then computed the typical time scales over which these processes occur. In particular, we have checked that our results are consistent with the decay rates obtained by Hawking and Moss~\cite{Hawking:1981fz}. We have also noticed that the ratio between the decay rates and the ratio between the decay probabilities exactly match in the steepest descent approximation.

We have also explained why our results are consistent with the stationary solutions~\cite{Starobinsky:1986fx, Linde:1993xx, Starobinsky:1994bd} of the Fokker-Planck equation, and we have derived the equilibration time scale to attain these equilibrium distributions. Since it typically corresponds to a gigantic number of \efolds, in the ``fall'' problem the relative populations  of the vacua is given by the almost symmetric result we derived in \Eq{eq:R:fall:SharpMax:appr} rather than by the highly asymmetric one arising in the stationary state.

Finally, we have studied tunnelling events in arbitrary potentials comprising several minima and maxima. We have shown that the integral structure of \Eq{r} guarantees that the tunnelling probabilities we derived satisfy the axiom of conditional probabilities. This is why our results for  the ``fall'' and for the ``escape'' problems are building blocks that can readily be put together to study more complex situations. Our work therefore provides a fairly comprehensive analysis of all possibilities for boundary crossing in single-field slow-roll models of inflation.

It would be interesting to investigate how these results generalise to multiple-field setups. Indeed, the dynamics of stochastic inflation  is highly sensitive on the number of fields it is driven by~\cite{Assadullahi:2016gkk, Vennin:2016wnk}. At the technical level, the ordinary differential equation~(\ref{original-p-eq}) becomes a partial differential equation for which there is no generic analytical solution such as \Eq{r}, which makes the problem more difficult to study. In a generic potential landscape, two given vacua are connected by several paths, and the continuous generalisation of the conditional probabilities~(\ref{p124}) may lead tunnelling probabilities, and their associated decay rates, to acquire new interesting properties. 
\acknowledgments
M.N.\ acknowledges financial support from the research council of University of Tehran. V.V. acknowledges funding from the European Union's Horizon 2020 research and innovation programme under the Marie Sk\l odowska-Curie grant agreement N${}^0$ 750491. H.A., V.V. and D.W. acknowledge support from the UK Science and Technology Facilities Council grant ST/N000668/1.
\appendix
\section{Steepest descent and asymptotic expansion}
\label{app:asymp-exp}
In the main body of this paper, we have encountered several integrals of the form
\bea
\int_0^X \ee^{-u(x)} \dd x\, ,
\eea
where the integral is dominated by its maximum at $x=0$, and $X$ is positive and large.  The strategy is to expand the integrand around $x=0$ and use the steepest descent approximation for the resulting Gaussian integral.  An example would be
\bea
\label{I-def}
I\left(a,b;X\right) = \int_0^X \ee^{-ax^2-bx^3} \dd x\, ,  \qquad a>0.
\eea
In fact, we can even have $b<0$, as long as the cubic term is small over the domain of integration.  To be more precise, we have to express this in terms of dimensionless quantities
\bea
\label{abX-approx}
aX^2\gg1 \qquad \text{and} \qquad a^3\gg b^2\, .
\eea
The reason why \eqref{abX-approx} is the right criterion for the validity of our approximation will be made clear as we proceed.

In most of our examples $u(x)$ does not diverge at infinity, so the integral is not convergent when $X=\infty$. This will be troublesome as we want to let $X=\infty$ as a first approximation. To avoid this difficulty, we introduce an extended version of $u$:
\bea
u_{\mathrm{ext}}(x) = \begin{cases}
u(x) & \mathrm{if}\ 0<x\leq X\, , \\
u(x) + \left[x-u(x)\right] e^{-1/(x-X)^2} & \mathrm{if}\ X<x\, .
\end{cases}
\eea
It has the advantage that the integral $\int_0^X e^{u_{\rm ext}(x)} \dd x$ is identical to $\int_0^X e^{u(x)} \dd x$, while it is convergent for $X\to\infty$. Furthermore, $u_{\rm ext}$ and all its derivatives coincide with those of $u$ at $x=X$. In the sequel we will drop the subscript ``ext'' and work exclusively with $u_{\rm ext}$. This will cause no problem, since our ultimate result is an expression for $\int_0^X e^{u(x)} \dd x$ in terms of $u$ and its derivatives at $x=0,X$; and none of these quantities change if we change $u$ to $u_{\mathrm{ext}}$.\footnote{Note that $u_{\rm ext}$ always depends on $X$, but $u$ may or may not depend on $X$. For example,  in the symmetry breaking example of \Sec{sec:fall:example}, $1/u = \lambda (x^2/X^2-1)^2 + \epsilon x^3/X^3 +\bar v_0$, so $u$ explicitly depends on $X$.}

We can now work out an asymptotic expansion for large but finite $X$.  This is similar to the asymptotic expansion of the error function [when $b=0$ in \Eq{I-def}].  One proceeds by successively integrating by parts:
\bea
\label{asymp-gen}
\int_X^\infty e^{-u}\dd x 
&= -\int_X^\infty \frac{\dd x}{\dd u} \dd \left(\ee^{-u}\right) 
= -x'e^{-u} \Big|_X^\infty + \int_X^\infty \ee^{-u} \dd x' \\
&= -x'\ee^{-u} \Big|_X^\infty - \int_X^\infty \frac{\dd x'}{\dd u} \dd\left(\ee^{-u}\right) 
= -x'\ee^{-u} \Big|_X^\infty - x''\ee^{-u} \Big|_X^\infty + \int_X^\infty \ee^{-u} \dd x'' \\
&= \ee^{-u(X)} \left[ x' + x'' + x''' + \ldots + x^{(n)} \right]_X + \int_X^\infty \ee^{-u} \dd x^{(n)},
\eea
where use has been made of the fact that $u(x)$ blows up at $x=\infty$. In this expression, $x'$ stands for $\dd x/\dd u$ and $x^{(n)}=\dd^n x/\dd u^n$. This is not a convergent series, but is an asymptotic one.\footnote{In an asymptotic series $s = \sum a_n$, we have $a_n = o(a_{n-1})$ and the remainder $R_N = s - \sum^N a_n$ is $o(a_N)$.  We have employed the little $o$ notation: $f=o(g)$, if $f/g\to0$ as $X\to\infty$.  Most of asymptotic series diverge, but that doesn't mean they are useless: the first few terms usually give a good approximation to the actual value before the inclusion of higher terms spoils the convergence.}  To prove this, note that $x^{(n)} / x^{(n-1)} = O(u^{-1})$,\footnote{Strictly speaking, this is not true for all functions $u(x)$, but it works for a wide class of functions including $u(X) = O(X^{n\neq0})$ --- provided that the non-extended version of $u$ is $X$-independent. It also works for the symmetry breaking example of \Sec{sec:fall:example}.} evaluated at $x=X$, tends to $0$ as $X\to\infty$.  Furthermore, since $x^{(n)} = O(u^{-n+1}) x'$ decays for large $u$, it takes its maximum on the integration interval $X<x<\infty$ around $x=X$.  Thus
\bea
\left\vert \int_X^\infty \ee^{-u} \dd x^{(n)} \right\vert 
&= \left\vert \int_{u(X)}^\infty \frac{\dd x^{(n)}}{\dd u} \ee^{-u} \dd u \right\vert
\leq \left[ \max_{x\geq X} \left\vert x^{(n+1)}\right\vert \right] \left\vert \int_{u(X)}^\infty \ee^{-u} \dd u \right\vert \\
&= O \left[ \left\vert x^{(n+1)}(X)\right\vert e^{-u(X)} \right] = o \left[ \left\vert x^{(n)}(X)\right\vert \ee^{-u(X)} \right]\, ,
\eea
so indeed the remainder decays faster than the last term.  

Inspection of $x^{(n)}$ reveals that
\bea
x^{(n)} = (-1)^{n+1} (2n-3)!! \frac{u''^{n-1}}{u'^{2n-1}} + \ldots,
\eea
with the other terms having a similar factorial-type $n$-dependence.  So $x^{(n)} \sim (2n)!! / u^{n-1}u'$.  This confirms the divergence of the series for any finite $u$ (or $X$).  But it also tells us at what order the series begins to diverge.  For a fixed $X$, this happens when the $x^{(n)} \sim x^{(n+1)}$, i.e., when $n\sim |u(X)| \gg1$ [for example, the first condition in~\eqref{abX-approx} guarantees that $u = aX^2+bX^3 \gg1$ in the case of \Eq{I-def}\footnote{Here and below, we are assuming that the sum of two terms is of the same order of magnitude as the greater of the two, hence $aX^2+bX^3 \gtrsim aX^2 \gg 1$.  This is correct, unless the two terms have opposite signs and there is a fine-tuning between the two terms such that their sum cancel out at leading order and the result becomes much smaller than any of them (for example, if $aX^2=10^3$ and $bX^3=-10^3+1$).  We assume that there is no such fine tuning.}].  So the first few terms cause no harm.

The derivatives $x^{(n)}=\dd ^nx/\dd u^n$ in \Eq{asymp-gen} must be rewritten in terms of derivatives $u^{(n)}=\dd^nu/\dd x^n$ of $u$ with respect to $x$.  For example, $x'=1/u'$, $x''=-u''/u'^3$, $x'''=(-u'u'''+3u''^2)/u'^5$, etc.  Thus for $u=ax^2+bx^3$, we arrive at
\bea
\int_X^\infty e^{{-ax^2-bx^3}}\dd x = \frac{e^{-aX^2-bX^3}}{2aX+3bX^2} \left[ 1 - \frac{2a+6bX}{(2aX+3bX^2)^2} + \ldots \right].
\eea
This implies that
\bea
\label{IX-vs-Iinf}
I(a,b;X) = I(a,b;\infty) - \frac{e^{-aX^2-bX^3}}{2aX+3bX^2} \left[ 1 - O \left( \frac{1}{aX^2+bX^3} \right) \right],
\eea
that is, the error in $I$ due to replacing $X$ with $\infty$ is exponentially small. This also shows that all we need is $u(X) \gg 1$, since the error is always given by $e^{-u}/u'|X$, regardless of the form of $u$. In particular, we don't need $aX^2 \gg bX^3$, and therefore $aX^2 \gg 1$ [the first condition of \eqref{abX-approx}] is so far sufficient for the validity of our approximation.

Our next task is to expand $u(x)$ near $x=0$ and turn the integral into a Gaussian.  The expansion of the exponent will be of the form
\bea
\label{u-taylor}
u(x) = \sum_{n=2}^\infty \frac{1}{n!} u^{(n)}(0) x^n,
\eea
where $u''(0)>0$ and we have assumed that the constant term ($n=0$) is already subtracted away, since it has only a multiplicative effect on $\int e^{-u}$.  We then expand the exponential function itself for all terms beyond the quadratic one.  For illustration, we proceed with the special case of $I(a,b;\infty)$ given by \Eq{I-def}, and obtain an expansion in terms of small $b^2/a^3$.  To do so we Taylor expand the cubic term, and using
\bea
\int_0^\infty x^k e^{-ax^2} \dd x= \frac{\Gamma({\frac{k+1}{2}})}{2a^{(k+1)/2}}\, ,
\eea
for $a>0$, we obtain
\bea
\int_0^\infty \ee^{-ax^2-bx^3} \dd x &= \int_0^\infty \sum_{n=0}^\infty \frac{(-bx^3)^n}{n!} \ee^{-ax^2} \\
&= \frac{1}{2\sqrt a} \sum_{n=0}^\infty (-1)^n \frac{\Gamma({\frac{3n+1}{2}})}{n!} \left( \frac{b^2}{a^3} \right)^{n/2}\, .
\eea
Of course, interchanging the sum and integral is not legitimate and that's why the resulting series is again divergent.  But this one too is an asymptotic series.  This time the onset of diverging terms is given by 
\bea
\frac{\Gamma({\frac{3n+4}{2}})}{\Gamma({\frac{3n+1}{2}})} \frac{n!}{(n+1)!} \left( \frac{b^2}{a^3} \right)^{1/2} \sim 1\, .
\eea
For large $n$ we find, using Stirling's formula, that this happens when $n\sim a^3/b^2 \gg 1$.  Thus, the first couple of terms improve the accuracy and only higher terms contribute to the divergence.

We observe that the quantity $b^2/a^3$ naturally appeared as our expansion parameter.  It is the second half of what we claimed in \Eq{abX-approx} was the criteria of validity of our approximations.  This can be justified as follows: In the absence of the cubic term, most of the contribution to the integral comes from $x<x_2$, where $x_2=1/\sqrt a$, beyond which the integrand is exponentially suppressed.  This makes it further clear why the first half of \eqref{abX-approx} is necessary (it says that $X\gg x_2$, so that the Gaussian result is a good approximation).  When the cubic term is present, its effect becomes comparable to that of the quadratic term at $x=x_3$, where $x_3=a/|b|$.  We don't want this term to spoil the Gaussian approximation.  Thus we need $x_3 \gg x_2$, that is, the expansion parameter is $b/a^{3/2} \ll 1$, which is precisely the second half of \eqref{abX-approx}.  If a quartic term $cx^4$ was present in $u$, then a second parameter of expansion, namely $c/a^2$, would appear from demanding $x_4\gg x_2$, where $x_4=\sqrt{a/|c|}$.

More generally, working with \Eq{u-taylor}, we require that none of the higher terms spoil the leading Gaussian picture.  This means that the conditions for the validity of our approximation are $X\gg x_2$, as well as $x_n \gg x_2$ for $n>2$, where $x_2$ and $x_n$ are given by
\bea
\frac{1}{2} \left\vert u''(0)\right\vert x_2^2 = 1 \qquad \mathrm{and} \qquad \frac{1}{2} \left\vert u''(0)\right\vert x_n^2 = \frac{1}{n!} \left\vert u^{(n)}(0) \right\vert x_n^n\, ,
\eea
respectively.  These can be rewritten as
\bea
\label{gen-approx-cond}
\gamma_2 \equiv \frac{2}{\left\vert u''(0)\right\vert X^2} \ll 1, \qquad
\gamma_{n>2} \equiv \left[ \frac{2}{\left\vert u''(0)\right\vert} \right]^{n/2} \frac{u^{(n)}(0)}{n!} \ll 1\, ,
\eea
and we identify $\gamma_n$ ($n\geq2$) as the parameters of the asymptotic expansion.

In summary, we have
\bea
\label{asymp-final-generic}
\int_0^X \ee^{-u(x)} \dd x = \frac12 \sqrt \frac{\pi}{u''(0)} \left[ 1 - \frac{\gamma_3}{\sqrt\pi} + O \left( \gamma_3^2,\gamma_4,\gamma_5,\ldots \right) \right] + O \left[ \frac{e^{-u(X)}}{u'(X)} \right]\, .
\eea
The conditions for the validity of this expansion is given by \eqref{gen-approx-cond} and $u(X)\gg1$.  In the special case of \Eq{I-def}, we find
\bea
\label{asymp-final}
\int_0^X e^{-ax^2-bx^3} \dd x = \frac12 \sqrt \frac{\pi}{a} \left[ 1 - \frac{1}{\sqrt\pi} \frac{b}{a^{3/2}} + O \left( \frac{b^2}{a^3} \right) + O \left( \frac{e^{-aX^2}}{\sqrt{aX^2}} \right) \right]\, ,
\eea
which is valid under the conditions~\eqref{abX-approx}, since $\gamma_2=1/aX^2$ and $\gamma_3=b/a^{3/2}$, and the exponentially small error term is in fact an upper bound on the error (it can be much smaller if $bX\gg a$).

\bibliographystyle{JHEP}
\bibliography{StochasticTunneling}
\end{document}